\documentclass[10pt, conference, letterpaper]{IEEEtran}
\IEEEoverridecommandlockouts
\pdfoutput=1
\usepackage{cite}
\usepackage{amsmath,amssymb,amsfonts}
\usepackage{graphicx}
\usepackage{textcomp}
\usepackage{xcolor}
\def\BibTeX{{\rm B\kern-.05em{\sc i\kern-.025em b}\kern-.08em
    T\kern-.1667em\lower.7ex\hbox{E}\kern-.125emX}}

\usepackage{algorithm, algpseudocode}
\newcommand {\abs}[1]{\left\vert #1 \right\vert}

\usepackage{array}
\usepackage{float}
\usepackage[labelformat=simple]{subcaption}

\usepackage{url}
\usepackage{hyperref}
\usepackage{flafter}
\usepackage{booktabs}
\usepackage{ifthen}
\usepackage{comment}
\usepackage{mdframed}

\begin{document}

\title{Cardinality Estimation in a Virtualized Network Device Using Online Machine Learning}

\author{
	\IEEEauthorblockN{Reuven Cohen}
	\IEEEauthorblockA{\textit{Department of Computer Science} \\
	\textit{Technion-Israel Institute of Technology}\\
	Haifa, Israel \\
	rcohen@cs.technion.ac.il}
	\and
	\IEEEauthorblockN{Yuval Nezri}
	\IEEEauthorblockA{\textit{Department of Computer Science} \\
	\textit{Technion}\\
	Haifa, Israel \\
	ynezri@cs.technion.ac.il}
}

\maketitle
\thispagestyle{plain}
\pagestyle{plain}

\begin{abstract}
	Cardinality estimation algorithms receive a stream of elements, with possible repetitions, and return the number of distinct elements in the stream. Such algorithms seek to minimize the required memory and CPU resource consumption at the price of inaccuracy in their output.  In computer networks, cardinality estimation algorithms are mainly used for counting the number of distinct flows, and they are divided into two categories: sketching algorithms and sampling algorithms. Sketching algorithms require the processing of all packets, and they are therefore usually implemented by dedicated hardware. Sampling algorithms do not require processing of all packets, but they are known for their inaccuracy. In this work we identify one of the major drawbacks of sampling-based cardinality estimation algorithms: their inability to adapt to changes in flow size distribution. To address this problem, we propose a new sampling-based adaptive cardinality estimation framework, which uses online machine learning. We evaluate our framework using real traffic traces, and show significantly better accuracy compared to the best known sampling-based algorithms, for the same fraction of processed packets.
\end{abstract}

\section{Introduction} \label{sec:intro}

	Network measurement plays an important role in the evolution of large scale networks. Traffic statistics, such as top-K \cite{mouratidis2006continuous} and heavy hitters \cite{ben2016heavy}, are important for crucial network management applications. Cardinality estimation is the problem of estimating the number of distinct elements in a data stream with repeated elements. In the field of network measurement, this problem is mainly related to counting the number of unique flows in an IP packet stream, where each flow consists of many packets that have a unique 5-tuple: source IP, destination IP, source port, destination port and Protocol. Finding the number of distinct flows is important for tracking the load imposed on a web server, detecting a potential Distributed Denial of Service (DDoS) attack, and discovering other network anomalies.
	
	Cardinality estimation algorithms are roughly divided into two categories: sampling algorithms and sketching algorithms. Sampling algorithms process only a subset of the stream, and use statistical analysis for estimating the cardinality of the entire stream. Such algorithms are very efficient in terms of their processing time and memory consumption. However, they have a relatively high error rate \cite{charikar2000towards}, especially when the stream has a skewed (power-law) distribution \cite{deolalikar2016extensive}.
	
	Sketching algorithms (HLL \cite{flajolet2007hyperloglog}, HLL++ \cite{heule2013hyperloglog}) are known to offer the best performance in terms of statistical accuracy and memory storage. However, their main disadvantage is that they must process all the packets in the data stream. For this reason they are very often implemented using dedicated hardware.
	
	Software network devices are gaining popularity due to the rise of two new network architecture concepts: Software Defined Networking (SDN)  and Network Function Virtualization (NFV). The transition from executing cardinality estimation algorithms by dedicated hardware to executing them by a general purpose CPU introduces new challenges. Even with a very small number of operations per packet, processing every stream packet by a general purpose CPU is inefficient and in many cases even impossible \cite{alipourfard2015re,alipourfard2018comparison}. Because it is usually impossible to process all the packets, sketching cannot be used. In such cases we must resort to sampling-based solutions.
	
	Many sampling cardinality estimation algorithms were developed for database query optimization. Since database records can represent any type of data, these algorithms make no prior assumption on the underlying data distribution. However, when network traffic is considered, information about flow size distribution can be used to improve the estimation precision. For example, the traffic generated by a DNS server usually includes many short flows, while the traffic generated by a video server has a small number of much longer flows. If the two servers produce the same amount of data, we expect the cardinality of the former type of traffic (DNS) to be greater than that of the latter (video). A problem with this approach is that the distribution of traffic flows is likely to change; e.g., due to dynamic server provisioning in the case of virtual environments, or due to a Network/Transport layer attack. We believe that dynamic adaptation of the cardinality estimation algorithms to changes in the flow size distribution can introduce significant improvement in their accuracy, and propose to use machine learning (ML) to obtain such adaptivity.
	
	In this paper we propose a novel sampling-based adaptive cardinality estimation framework, which incorporates online ML algorithms to adapt to changes in flow size distribution. We describe the framework and its parameters, and then discuss the selection of an online ML algorithm and its features. We evaluate our framework using real traffic traces and show its accuracy improvement over the best known sampling algorithms.
	
	The rest of the paper is organized as follows. Section \ref{sec:related} discusses related work. Section \ref{sec:perliminaries} describes the basic concepts of sampling-based cardinality estimation and online learning. Section \ref{sec:framework} presents the proposed framework for adaptive sampling cardinality estimation. Section \ref{sec:evaluation} presents an extensive evaluation of our framework and compares it with the best-known sampling-based cardinality estimation algorithms. Section \ref{sec:conclusions} concludes the paper.
	
\section{Related Work} \label{sec:related}

	Many works address the cardinality estimation problem and propose efficient algorithms to solve it. A detailed survey of the problem and its various solutions is presented in \cite{gibbons2016distinct}. 
	
	In this section we first discuss sampling-based solutions and cardinality estimation algorithms whose estimation process is adaptive. Then, we discuss works that analyze the efficiency and accuracy of software-based sketches. Many of the works we discuss here refer to the more general problem of estimating the flow size distribution, namely, the number of flows that contain a specific number of packets. This problem can be easily reduced to cardinality estimation.
	
	In \cite{haas1998estimating}, Hass et al.\ present a family of cardinality estimators based on the generalized jackknife technique. They present an ``unsmoothed first-order jackknife estimator'' (UJ1) and its ``smoothed'' version (SJ1). They also present an ``unsmoothed second-order jackknife estimator'' (UJ2) and its ``smoothed'' and ``stabilized'' versions (SJ2 and UJ2A). 
	
	In \cite{charikar2000towards}, Charikar et al.\ present a lower bound on the error of sampling-based estimators. This lower bound implies that one should process almost the entire stream in order to guarantee good estimation error over all possible inputs. Then, they present Guaranteed Error Estimator (GEE) and Adaptive Estimator (AE). GEE has an optimal error, which matches their proved bound. AE is a refinement of GEE, which adapts to the input distributions and obtains reduced error over low-skewed data.
	
	Recent work by Deolalikar et al.\ \cite{deolalikar2016extensive} conducts an extensive comparison of 11 sampling-based cardinality estimators and tests their accuracy over power-law distributed data with different skew parameters. GEE, AE and UJ2A \cite{haas1998estimating} are found to be the most accurate estimators over a variety of data distributions. We shall refer to these three algorithms in the evaluation of our framework (Section \ref{sec:evaluation}).
	
	In \cite{duffield2003estimating}, Duffield et al.\ take advantage of the SYN flag in the TCP header to obtain additional information about flow size distribution. This method requires that most of the packets in the stream be TCP. This requirement is not trivial in real traffic, due to the increasing popularity of the new UDP-based transport layer protocols \cite{hamilton2016quic}. Our framework can also use the number of SYN packets to improve the estimation accuracy, but it does not rely only on protocol specific information. 
	
	In \cite{cohen2017cardinality}, a novel hybrid approach is presented. It combines Good-Turing frequency estimation \cite{gale1995good}, a sampling-based solution, and the HyperLogLog algorithm \cite{flajolet2007hyperloglog}. This solution benefits from the computational efficiency of sampling and from the memory efficiency of sketching, but its accuracy is bound to that of sampling-based algorithms.
	
	Alipourfard et al.\ \cite{alipourfard2015re} show that in a software switch (OVS \cite{pfaff2015design}), calculating stream statistics using a simple hash table achieves better throughput and latency. They claim that the main reason is the trade-off between memory and CPU consumption imposed by sketching algorithms. Maintaining a sketch usually requires high CPU consumption, mainly for hashing each packet multiple times. However, modern servers have significantly improved cache size and efficiency. Hence, they do not really benefit from the reduced memory usage. CPU consumption is observed in their later work \cite{alipourfard2018comparison} as the new bottleneck resource.

\section{Preliminaries} \label{sec:perliminaries}
	
	As indicated in the previous sections, a software device should not use sketching. On the other hand, sampling-based algorithms are known to have relatively low accuracy. To address this trafe-off, the framework proposed in this paper combines sampling with online machine learning (ML). We start with a short discussion of each of these concepts.
	
	\subsection{Estimating Stream Cardinality from a Sample} \label{subsec:sampling}
	
		Before we discuss the details of estimating stream cardinality from a sample, we formally define the problem. Our notations are summarized in Table \ref{table:notations}. Let $S$ be a stream of $N$ packets belonging to a certain number of flows. Each flow typically contains many packets, and $D$ denotes the number of unique flows in $S$. Consider a sample $s$ of $n$ packets, taken randomly from $S$, and let $q=\frac{n}{N}$ be the sampling rate. Let $d$ represent the number of unique elements in $s$. Sampling based cardinality estimation is the process of providing an estimate $\hat{D}$ of $D$ given only $s$ and $q$.
		
		\begin{table}
			\centering
			\begin{tabular}{|c|l|} \hline
				\textbf{Symbol} & \textbf{Meaning}  \\\hline
				$S_i$ & a batch of packets (part of the stream)  \\\hline
				$N_i$ & size of $S_i$ (number of packets in $S_i$)\\\hline
				$D_i$ & cardinality of $S_i$ (number of flows in $S_i$)\\\hline
				$s_i$ & a sample of packets taken from $S_i$ \\\hline
				$n_i$ & size of $s_i$ (number of packets in $s_i$) \\\hline
				$q_i$ & sampling rate $q_i = n_i/N_i$ \\\hline
				$d_i$ & cardinality of $s_i$ (number of flows in $s_i$)\\\hline
				$\hat{D}_i$ & estimation of $D_i$ \\\hline
				$f_i^j$ & number of flows that appear exactly $j$ times\\& in $s_i$ (have exactly $j$ packets in $s_i$) \\\hline
			\end{tabular}
			\caption{Notations} 
			\label{table:notations}
		\end{table}
		
		A typical workflow of a sampling cardinality estimation algorithm over network traffic is as follows. The entire traffic stream $S$ is divided into batches $[S_1, S_2, ..., S_i, ...]$, and each batch $S_i$ is analyzed separately. A sample $s_i$ of $n_i$ packets is collected from $S_i$ according to a specific sampling rate $q$, using some sampling method. While several sampling methods have been proposed in the past \cite{einziger2017constant}, in this paper we consider random sampling, where all packets are sampled with the same probability. After a sample of packets is collected from the batch, various statistical properties are calculated from it, and an estimation of the flow cardinality is produced. 
		
		A commonly used statistical property is $f_i^j$, namely, the number of flows that appear exactly $j$ times in $s_i$. A special case is $f_i^0$, which represents the number of flows that appear in $S_i$ but not in $s_i$. Since $D_i=d_i+f_i^0$,  estimating $D_i$ can be reduced to estimating $f_i^0$, i.e.\, the number of ``hidden'' flows. Various works discuss the relation between $f_i^0$ and $f_i^j$ values. For example, the Good-Turing frequency estimation technique \cite{good1953population} claims that $\frac{f_i^1}{n}$ is a consistent estimator of $f_i^0$.

	\subsection{Online Machine Learning}
	
		Describing the details of ML is beyond the scope of this work. We therefore describe only the standard ML procedures we use. We use Scikit-Learn's abstraction of ML algorithms \cite{pedregosa2011scikit} throughout the paper. According to this abstraction, a supervised ML algorithm includes two basic operations:
		
		\begin{itemize}
			\item $fit(training\_examples, labels)$ -- The fit operation receives a set of training examples and their matching labels, and returns a ``trained ML model'', which is capable of delivering predictions from unlabeled examples. Each example consists of a set of features, where every feature is a property of the data that contributes to the estimation. For example, to train a ML model to predict housing prices, the features could be the size of the house, the number of rooms and the neighborhood. The label is the desired output value, i.e., the actual price of each house. For clarification, we use the term ``ML model'' to describe an instance of an ML algorithm after it is provided with training data.  
			
			\item $predict(example)$ -- After an ML model is trained using the \emph{fit()} operation, the predict operation is used to obtain an estimated value given only a set of features. In our housing prices example, \emph{predict()} is provided with information about a specific house: its size, number of rooms and neighborhood, and returns its price estimation.
		\end{itemize}
		
		A typical ML algorithm usually receives the entire set of training examples in advance. However, processing the entire training dataset at once is not feasible when analyzing large data streams that cannot entirely fit into the memory, or when the training examples are only gradually available. To address this problem, online ML algorithms have been developed. Such algorithms use the following \emph{partial\_fit()} operation, instead of \emph{fit()}:
		\begin{itemize}
			\item $partial\_fit(training\_example, label)$ -- This operation is similar to \emph{fit()}, except that it trains the model over a single or a small number of labeled training examples. While \emph{fit()} is executed only once, at the beginning of the model's lifespan, \emph{partial\_fit()} can be invoked many times.
		\end{itemize}
		
		In contrast to \emph{fit()}, \emph{partial\_fit()} allows the model to adapt over time. For example, in the context of housing prices, very often there are not enough examples of houses and their true market price. With \emph{partial\_fit()}, each sold house can be added to the model as a new example. If during the model's lifespan there is a trend to prefer houses in different neighborhoods, such a trend is likely to be reflected by the new training examples, and the model is likely to be more accurate compared to a model that receives all its training examples in advance.

\section{The Proposed Framework for Sampling-Based Adaptive Cardinality Estimation} \label{sec:framework}

	\subsection{Framework Description}

		\begin{figure*}[!bt]
			\centering
			\begin{subfigure}{.45\textwidth}
				\centering
				\includegraphics[width=1\linewidth]{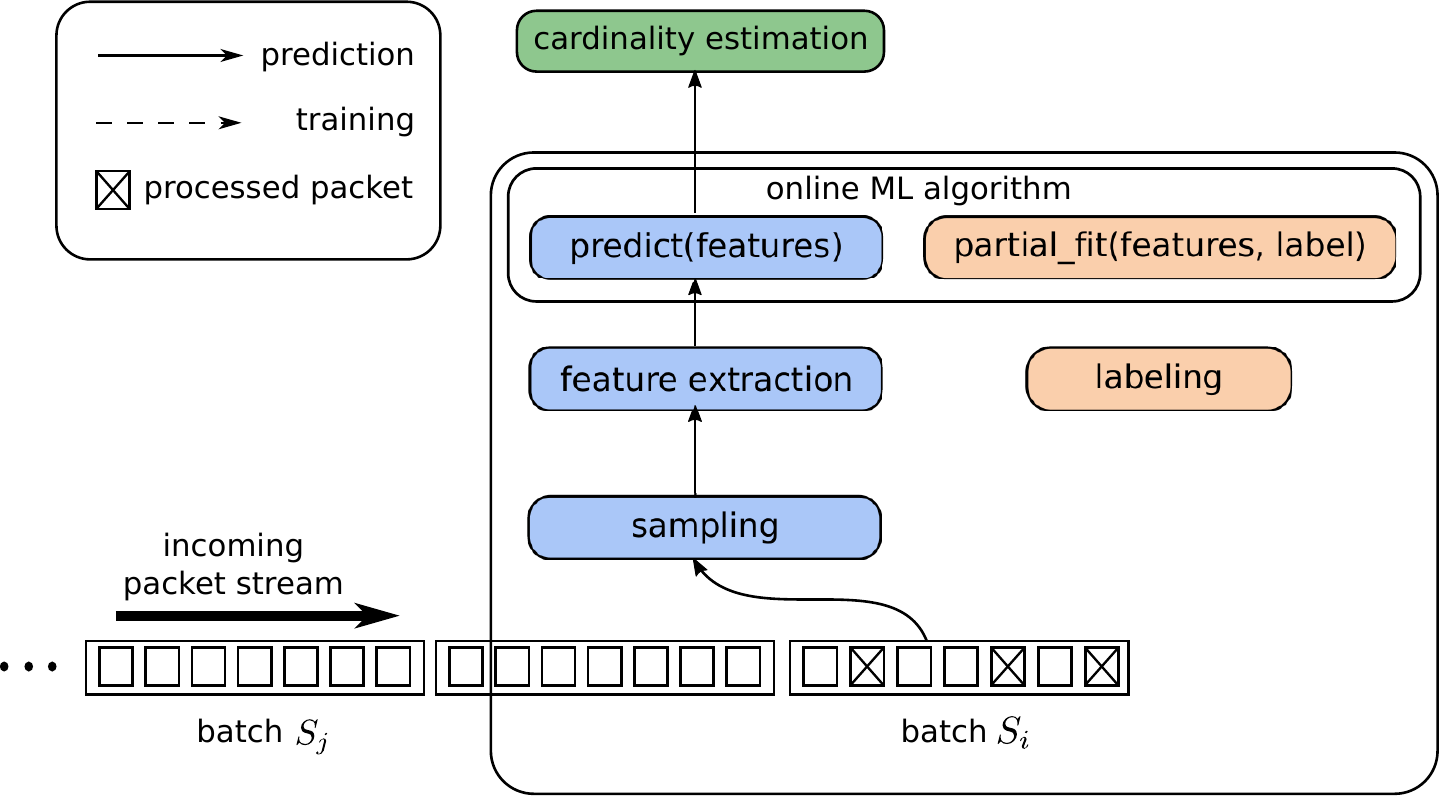}
				\caption{Prediction on batch $S_i$}
				\label{fig:prediction}
			\end{subfigure}
			\hfill
			\begin{subfigure}{.45\textwidth}
				\centering
				\includegraphics[width=1\linewidth]{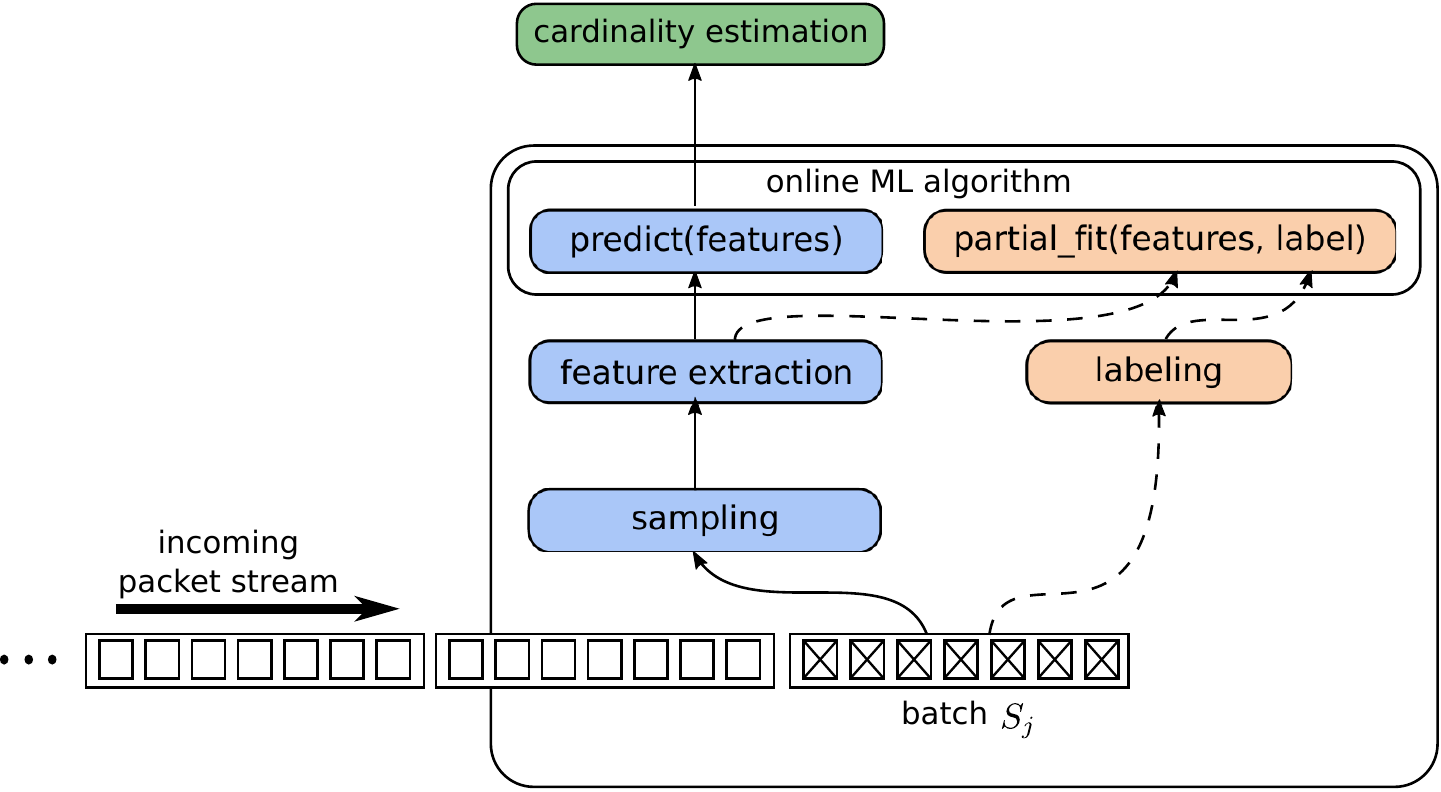}
				\caption{Prediction and training on batch $S_j$}
				\label{fig:training}
			\end{subfigure}
			\caption{The proposed adaptive cardinality estimation framework}
			\label{fig:framework}
		\end{figure*}
		
		Our sampling-based adaptive cardinality estimation framework uses a novel approach: combining sampling and online learning. Figure \ref{fig:framework} describes the processing of a packet stream in detail. The stream is divided into batches of packets. Each batch is sampled, and selected features are extracted from the samples. The features are then passed to the $predict()$ operation of the online ML algorithm, which returns an estimation of the batch's cardinality (Figure \ref{fig:prediction}).
		
		Once every $training\_rate$ batches, the entire batch is sent for training (Figure \ref{fig:training}). In the training phase, $partial\_fit()$ is provided with the batch's set of features and its cardinality. The exact cardinality of a batch can be calculated using a simple hash table. Since the focus of this work is on the framework itself, we intentionally do not specify which online ML algorithm is used. As shown in Section \ref{subsec:ol_alg}, different algorithms may be suitable for different framework use cases. Procedure \ref{alg:model} presents the pseudo code for processing a single batch as described above. 
		
		\begin{algorithm}
			\floatname{algorithm}{Procedure}
			\caption{Processing of a Batch}
			\label{alg:model}
			\begin{algorithmic}[1]
				\State $batch\_counter \gets 0$ \label{op:counter}
				\State $ml\_model.init()$ \label{op:init}
				
				\Function{process\_batch}{\emph{batch}} 
				\State $sample \gets sample(batch, sampling\_rate)$ \label{op:sample}
				\State $features \gets extract\_features(sample)$  \label{op:features_extract}
				
				\If {$is\_training\_batch(batch\_counter,training\_rate)$} \label{op:train_phase} \Comment{executed concurrently}
				\State $label \gets calculate\_label(batch)$ \label{op:label}
				\State $ml\_model.partial\_fit(features, label)$ \label{op:partial_fit}
				\EndIf \label{op:end_phase}
				
				\State $batch\_counter \gets batch\_counter + 1$
				\State \textbf{return} $ml\_model.predict(features)$ \label{op:predict}
				
				\EndFunction
			\end{algorithmic}
		\end{algorithm}	
		
		In steps \ref{op:counter} and \ref{op:init}, the batch counter and the online ML model are initiated. In steps \ref{op:sample} and \ref{op:features_extract}, each batch is sampled and selected features are extracted from it. In step \ref{op:train_phase}, batches are selected for training according to the chosen \emph{training\_rate} hyperparameter. In step \ref{op:label}, the true cardinality of training batches is calculated. In step \ref{op:partial_fit}, the training batch's feature set and true cardinality are input to the model's \emph{partial\_fit} operation. Finally, in step \ref{op:predict} an estimation of the batch's cardinality is returned. Since the training operations (steps \ref{op:train_phase}--\ref{op:end_phase}) are resource and time consuming, they can be executed in the background without delaying the estimation.
		
	\subsection{Feature Selection} \label{subsec: feature_select}
	
		Choosing the right set of features is crucial to the accuracy of an ML algorithm. A good feature should be both informative and independent: an informative feature shows high correlation to the target value, while an independent feature shows low correlation to the other features.
		
\begin{figure*}[!t]
			\centering
			\includegraphics[width=.95\linewidth]{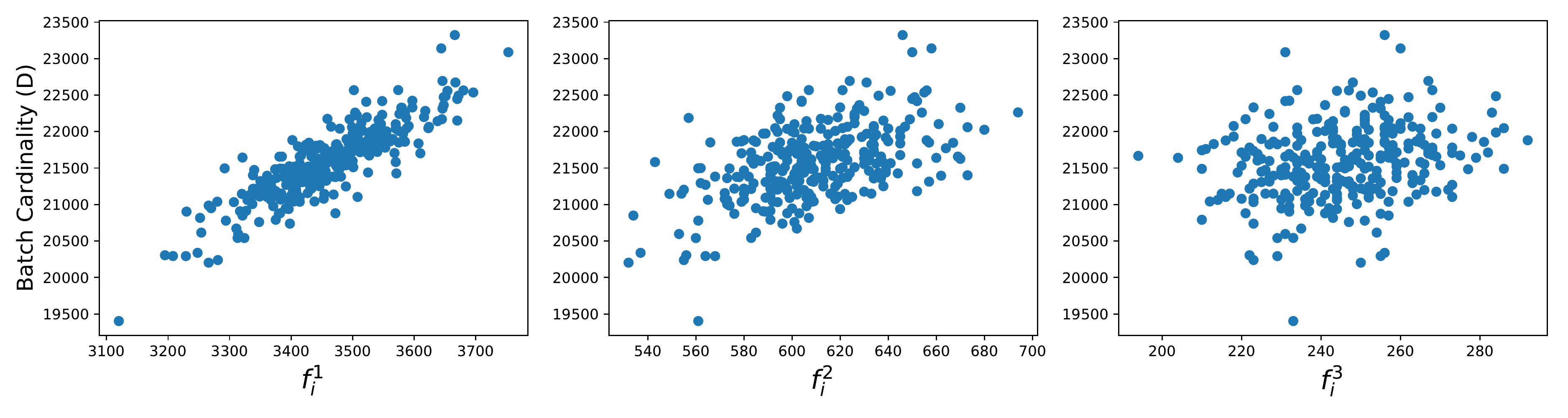}
			\caption{The correlation between $f_i^1$, $f_i^2$, $f_i^3$ and flow cardinality ($q=0.1$). We can see strong linear correlation between $f_i^1$ and the cardinality, and almost no correlation between $f_i^3$ and the cardinality}
			\label{fig:f_i_vs_card}
		\end{figure*}

		The first ``suspects'' for such features are the $f_i^j$ values, which are also used by statistical sampling-based algorithms. Recall that for a given batch $S_i$, $f_i^j$ is the number of flows that are represented by exactly $j$ packets in $s_i$. Figure \ref{fig:f_i_vs_card} shows the relationship between $f_i^1$, $f_i^2$, $f_i^3$ and the cardinality of $S_i$ ($D_i$). This graph is obtained using 300 batches, each of 100,000 packets, taken from CAIDA-2016 traffic trace\footnote{Section \ref{sec:evaluation} contains a detailed description of the CAIDA-2016 trace.}. Batches are sampled with $q=0.1$. We can see a strong linear correlation between $f_i^1$ and the cardinality. We can also learn that the correlation becomes weaker as $j$ increases.
		
\begin{figure*}[!bt]
			\centering
			\begin{subfigure}{.47\textwidth}
				\centering
				\includegraphics[width=.95\linewidth]{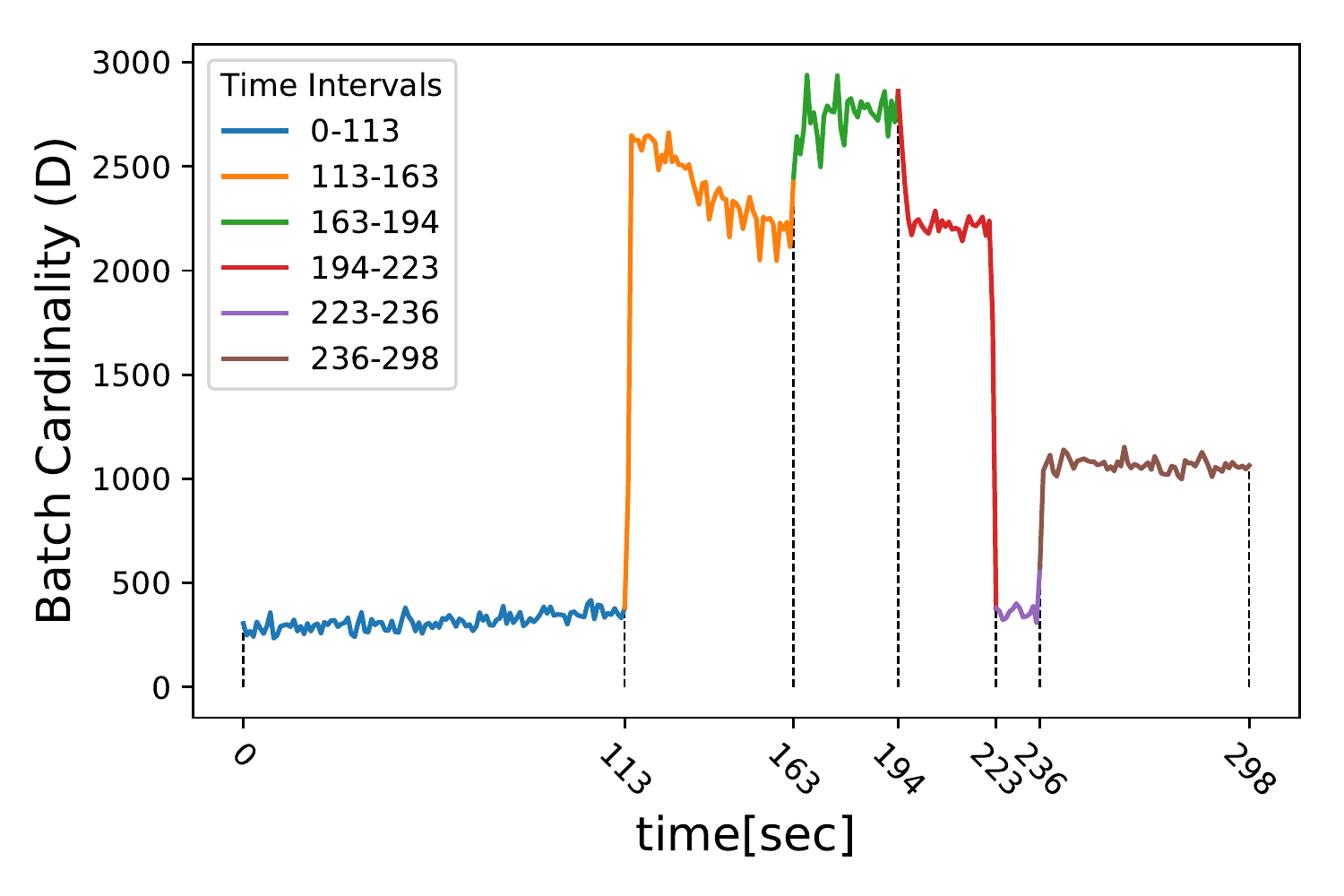}
				\caption{Batch cardinality as a function of time}
				\label{fig:batch_card_vs_time}
			\end{subfigure}
			\hfill
			\begin{subfigure}{.47\textwidth}
				\centering
				\includegraphics[width=.95\linewidth]{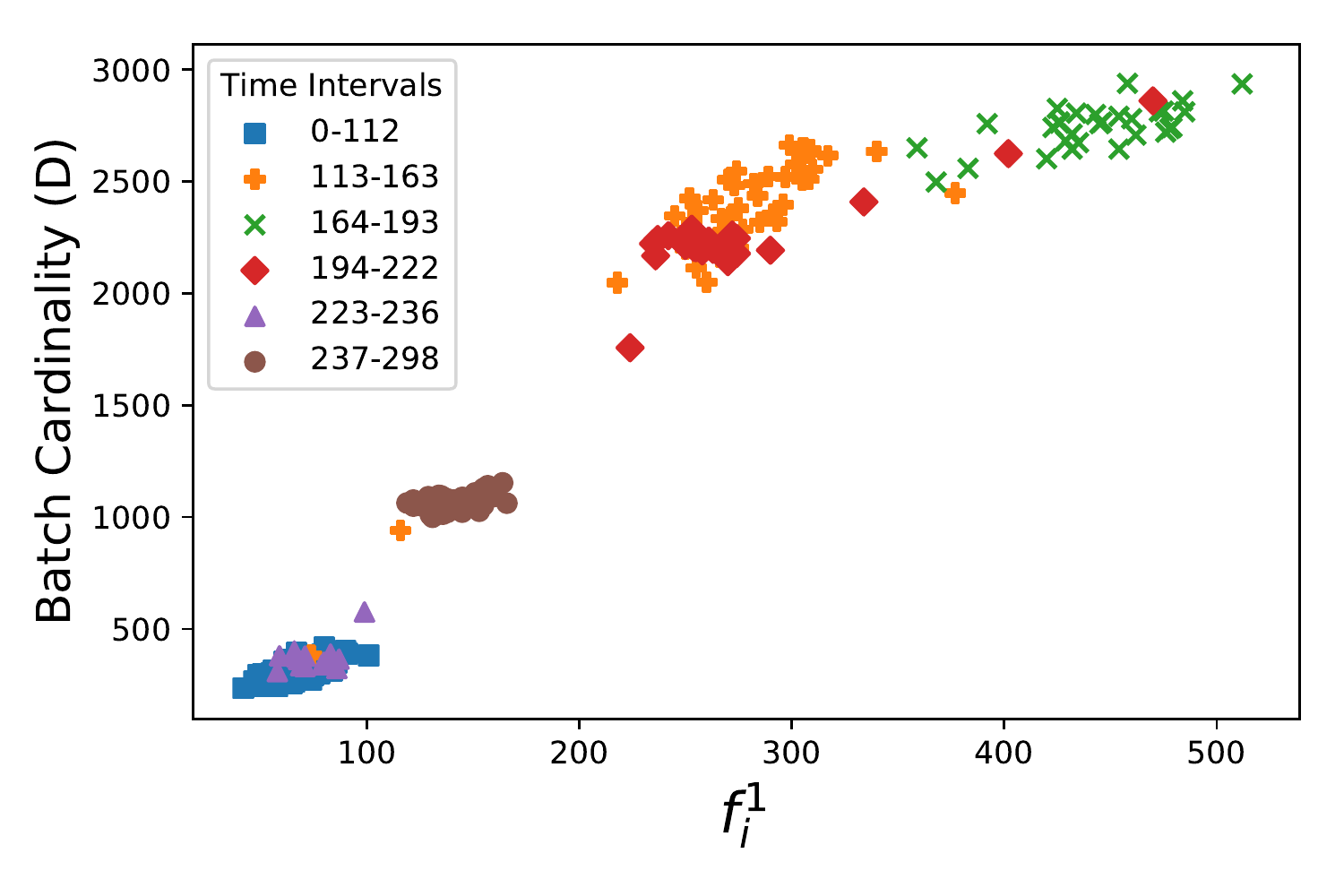}
				\caption{Batch cardinality as a function of $f_i^1$}
				\label{fig:f1_vs_card_darpa}
			\end{subfigure}

			\caption{The value of $f_i^1$ at different stages of the attack for DARPA-DDoS ($q=0.1$)}
			\label{fig:freq_feature}
		\end{figure*}

		When we test the correlation between $f_i^1$ and the cardinality for other traces, we see that the linear correlation is usually conserved, but the slope and intercept vary, since they depend on the flow size distribution. For example, Figure \ref{fig:batch_card_vs_time} shows batch cardinality over time for the DARPA-DDoS trace \cite{darpaddos}. This trace contains background traffic and a SYN flood DDoS attack on one target host. Each batch contains traffic collected during one second, and the average batch size is 11,228 packets. Batch no. 219 is removed from the trace since it shows abnormal behavior, which is irrelevant for the purpose of this example. We sample this trace with $q=0.1$.
		
		We can clearly divide the attack into 6 different stages, each beginning in one of the following time steps: $t=0, 113, 163, 194, 223, 236$. Figure \ref{fig:f1_vs_card_darpa} shows the cardinality of the batch as a function of $f^1$, and Figure \ref{table:darpa_pearson} describes the slope, intercept and Pearson correlation coefficient obtained by running a linear regression on each interval separately.
		
		The Pearson correlation coefficient is a number between -1 and 1 that indicates the extent to which two variables are linearly related, where -1 indicates a perfect negative linear relation, 0 indicates no linear relation and 1 indicates a perfect positive linear relation. The Pearson correlation coefficient values in Figure \ref{table:darpa_pearson} indicate a strong linear relation between $f_i^1$ and cardinality in most stages. The slope and intercept values indicate significant differences in the linear relation of different stages. We expect our online ML algorithm to identify these changes and to adjust its cardinality estimation accordingly.

		\begin{figure}
			\centering
			\small
			\begin{tabular}{cccc}
\toprule
   Interval &  Slope & Intercept &  Pearson Coefficient \\
\midrule
   (0, 112) &   2.49 &    144.17 &                 0.68 \\
 (113, 163) &   7.57 &     261.5 &                 0.91 \\
 (164, 193) &   1.84 &  1,928.54 &                 0.67 \\
 (194, 222) &   2.98 &   1,431.4 &                 0.87 \\
 (223, 236) &   2.99 &    149.88 &                 0.54 \\
 (237, 298) &   1.21 &    895.08 &                 0.43 \\
\bottomrule
\end{tabular}
		
			\caption{Linear regression parameters of the different stages of the attack for DARPA-DDoS ($q=0.1$)}
			\label{table:darpa_pearson}
		\end{figure}
		
		While $f_i^j$, and in particular $f_i^1$, seem promising as features from which to learn about the cardinality, we can extract even more valuable information from a sample of IP packets. For example, sending a large amount of data usually involves long flows of relatively big packets. In the presence of such flows we expect the cardinality of the stream to decrease. Hence, we expect to find a negative correlation between the sample's average packet length and the stream's cardinality. 
		
\begin{figure*}[!bt]
			\centering
			\begin{subfigure}{.49\textwidth}
				\centering
				\includegraphics[width=.95\linewidth]{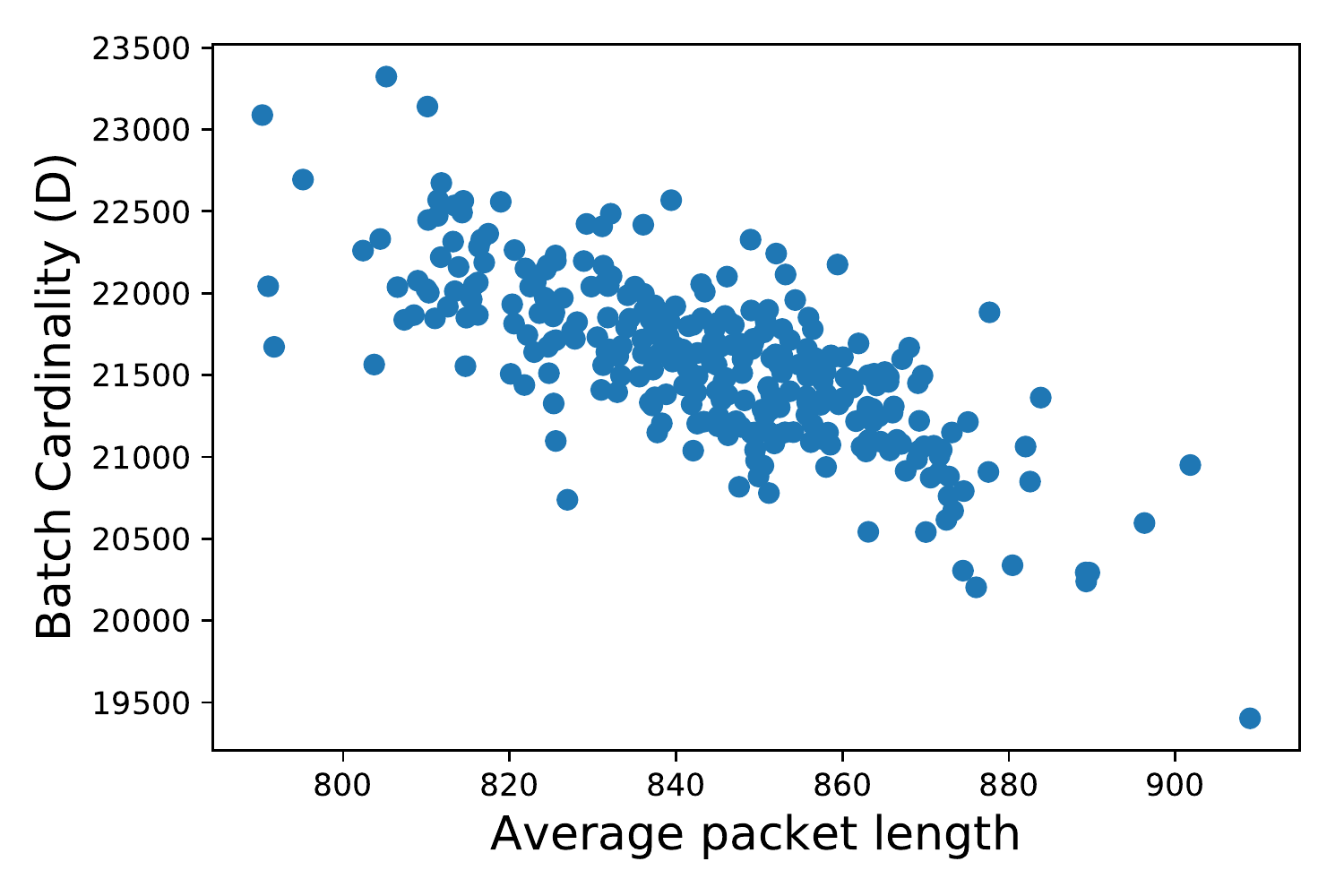}
				\captionsetup{justification=centering}
				\caption{Batch cardinality as a function \\ of $avg\_pkt\_len$}
				\label{fig:len_vs_card_backbone}
			\end{subfigure}
			\begin{subfigure}{.49\textwidth}
				\centering
				\includegraphics[width=.95\linewidth]{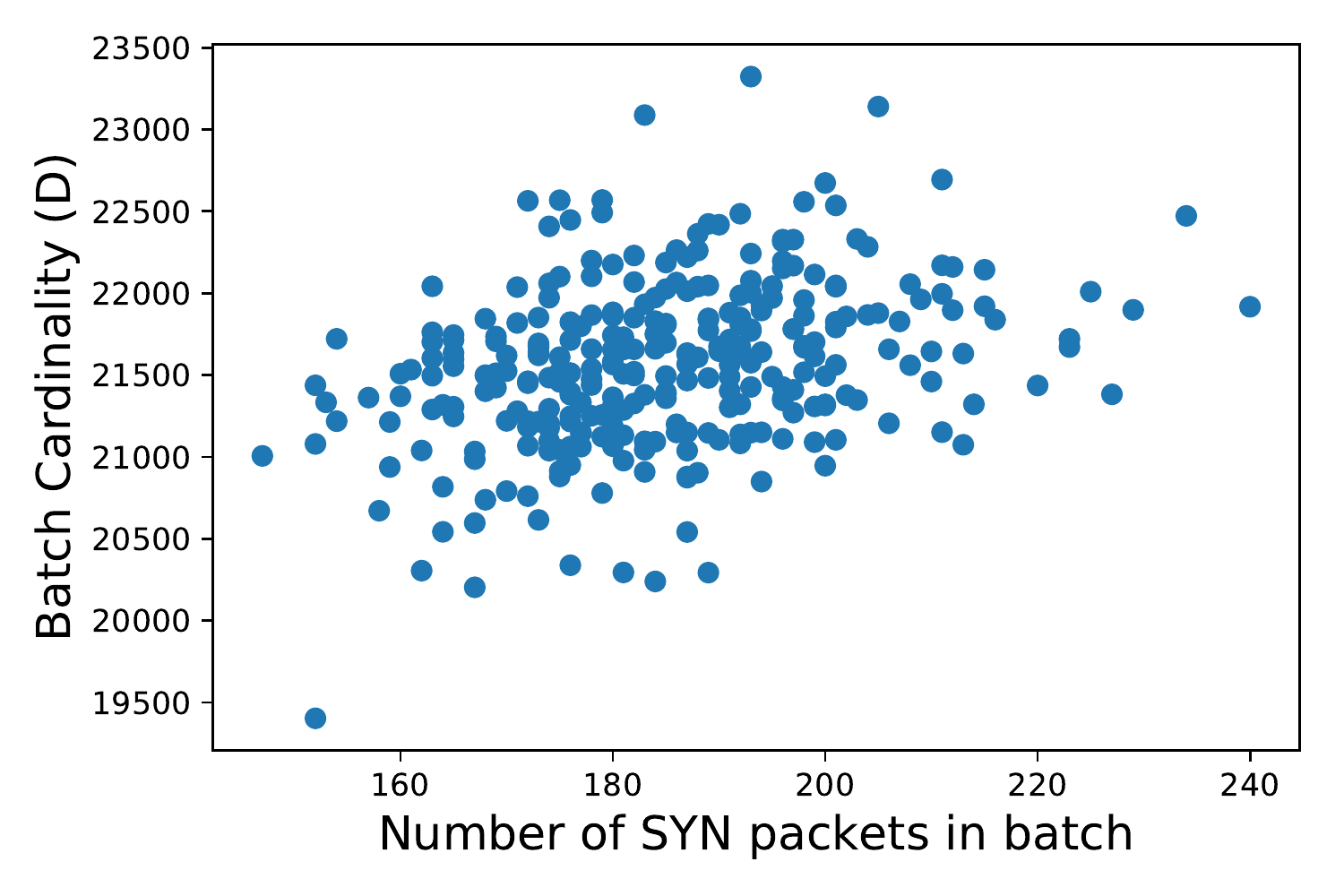}
				\captionsetup{justification=centering}
				\caption{Batch cardinality as a function \\ of $syn\_count$}
				\label{fig:syn_vs_card_backbone}
			\end{subfigure}
			\caption{The CAIDA-2016 trace with specific features ($q=0.1$)}
			\label{fig:tcp_feature}
		\end{figure*}

		Moreover, when most of the traffic is TCP, we expect to find a correlation between the number of SYN packets in the sample and the stream cardinality, because each SYN packet usually represents a single TCP connection. Figure \ref{fig:len_vs_card_backbone} shows the correlation between the average length of a packet in the sample (\emph{avg\_pkt\_len}) and the batch cardinality, and Figure \ref{fig:syn_vs_card_backbone} shows the correlation between the number of SYN packets in the sample (\emph{syn\_count}) and the batch cardinality. Both figures are from the CAIDA-2016 trace with $q=0.1$.
		
		In our study we choose $f_i^1$, $f_i^2$, $f_i^3$, $avg\_pkt\_len$ and $syn\_count$ as possible features, since they demonstrate good correlation to $D$ and they can be extracted from the sample. However, any such property can be added as a feature to our framework.
		
	\subsection{Choosing an Online ML Algorithm} \label{subsec:ol_alg}

		Since all the features we consider seem to present linear correlation with $D$, our first choice is a linear algorithm. Linear algorithms operate on assumption that the target value $D$ is a linear combination of the features. Let $\vec{x}$ be the vector of features and $\vec{w}$ be a vector of linear coefficients, also known as the weights vector. In each estimation phase, the $predict(\vec{x})$ operation returns $\hat{D} = \vec{w} \cdot \vec{x}$. In each training phase, the $partial\_fit(\vec{x}, D)$ operation updates $\vec{w}$ according an update rule that aims to minimize a predefined notion of estimation error. This notion of prediction error is commonly called the loss function. The way the update rule and loss function are defined determines different properties of the online ML algorithm, such as:

		\begin{itemize}
			\item Aggressiveness -- The intensity of updates in response to loss. 
			\item Forgetting Rate -- The effect of new training batches on the learning, compared to the effect of previous training batches.
			\item Outlier Sensitivity -- How the algorithm responds to training batches that introduce extreme loss. 
		\end{itemize}
		
		As shown in Figure \ref{fig:f1_vs_card_darpa}, the linear relation between a single feature and flow cardinality may frequently change. These changes may be attributed, for example, to different stages of a DDoS attack. Thus, the chosen algorithm must be able to react rapidly to such changes. In Section \ref{sec:evaluation} we compare 3 popular linear regression ML algorithms: Stochastic Gradient Descent (SGD) \cite{kiefer1952stochastic}, Recursive Least Squares (RLS) \cite{grant1987recursive} and Passive Aggressive (PA) \cite{crammer2006online}. 
		
		We describe PA and RLS since we found them to be the best performing algorithms in our experiments. We also include SGD because it is the foundation for a wide family of online ML algorithms. We experimented with more advanced SGD based algorithms (RMSProp {\cite{tieleman2012lecture}}, ADAgrad {\cite{duchi2011adaptive}} and ADAM {\cite{kingma2014adam}}), but they are excluded from this work since they did not show significant improvement over the basic version (SGD).
	
		\subsection{Rate Parameters} \label{subsec:rate_params}
			
			Obtaining timely and accurate estimation using low rate sampling is not trivial. Our framework has several rate parameters, which together have a crucial impact on the trade-off between accuracy and computational cost. Table \ref{table:rate_params} summarizes these parameters.
			
			\begin{table}[!tb]
				\centering 
				\begin{tabular}{|c|l|} \hline
					\textbf{Notation} & \multicolumn{1}{c|}{\textbf{Description}}  \\\hline
					\emph{packet\_rate} & incoming packet rate \\\hline
					\emph{batch\_size} & size of each processed batch \\\hline
					\emph{sampling\_rate} & fraction of packets sampled from each batch \\\hline
					\emph{estimation\_rate} & number of cardinality estimations\\
					& executed per second  \\\hline
					\emph{training\_rate} & number of batches between subsequent\\
					& training phases \\\hline
					\emph{sample\_size} & the size of a single sample \\\hline
					\emph{update\_rate} & number of training phases per second \\\hline
					\emph{effective\_sampling\_rate} & the total fraction of packets being processed,\\
					& including training phase \\\hline
				\end{tabular}
				\caption{Main Parameters of the Proposed Framework} 
				\label{table:rate_params}
			\end{table}
			
			The traffic stream can be divided into batches using two different approaches. In the first approach, \emph{batch\_size} is fixed, and a prediction phase starts after \emph{batch\_size} packets are processed. This approach is used when the estimation is not time critical, and does not have to be provided during fixed time intervals. In the second approach, \emph{estimation\_rate} is fixed, and a prediction phase is invoked exactly every $1/\emph{estimation\_rate}$ seconds. This approach is used when the time between consecutive estimations must be fixed.
			
			When \emph{batch\_size} is fixed, \emph{estimation\_rate} is determined by the $\emph{packet\_rate}/\emph{batch\_size}$ ratio. In this case, when \emph{packet\_rate} decreases, \emph{estimation\_rate} decreases as well. Eventually, if \emph{packet\_rate} is too low, we might end up with an \emph{estimation\_rate} that is not sufficiently high for our measurement application. For example, if the measurement application is DDoS detection, a measurement provided every 300 seconds might not be sufficiently frequent.
			
			In addition, \emph{estimation\_rate} has an impact on the framework's \emph{update\_rate} since \emph{update\_rate} is determined by $\emph{estimation\_rate} \cdot \emph{training\_rate}$. When \emph{estimation\_rate} is too low, the model may react too slowly to changes in the flow size distribution.
			
			When \emph{estimation\_rate} is fixed, \emph{batch\_size} is determined by the $\emph{packet\_rate}/\emph{estimation\_rate}$ ratio. In this case, when \emph{packet\_rate} decreases, \emph{batch\_size}  and \emph{sample\_size} decrease as well, and we might end up with a statistically insufficient \emph{sample\_size}.

\section{Performance Analysis} \label{sec:evaluation}

	We evaluated our new framework using 4 different traffic traces. In this section we report our findings for all these traces. We compare the framework's accuracy to that of statistical sampling-based algorithms, and analyze the effect of the various online ML algorithms and framework parameters on the accuracy. We also discuss the trade-offs imposed by the various framework parameters.  
	
	\subsection{Accuracy Metrics} \label{subsec:metrics}

		We use the following metrics in our evaluation:
		\begin{itemize}
			\item Root Mean Squared Error (RMSE), defined as $\sqrt{\frac{1}{n} \sum_{i=1}^{n} \left( D_i-\hat{D}_i \right) ^ 2}$. RMSE indicates the average prediction error in units of number of flows. Since the errors are squared before they are averaged, RMSE gives a higher weight to large errors, and penalizes high variance in the error distribution.
			
			\item Mean Absolute error (MAE), defined as $\frac{1}{n} \sum_{i=1}^{n}|D_i-\hat{D}_i|$. MAE also indicates the average prediction error in units of number of flows. As opposed to RMSE, it takes only the average error into account and is not affected by the error variance.
			
			\item Mean Absolute Percentage Error (MAPE), defined as $\frac{100}{n}\sum_{i=1}^{n}\abs{\frac{D_i-\hat{D}_i}{D_i}}$. MAPE indicates the average error percentage. It is biased towards underestimations, since for such estimations the percentage error cannot exceed 100\%, while for overestimations there is no upper limit. We use MAPE since it allows us to compare the accuracy over different traces and batch sizes even when batches have significant differences in their cardinality.
			
			\item Max Absolute Error (MAXAE), defined as $\max(|D_i-\hat{D}_i|)$. It indicates the maximum estimation error, and allows us to analyze the error in extreme cases.
			
		\end{itemize}

	\subsection{Effective Sampling Rate} \label{subsec:effective}

		Both the proposed framework and statistical sampling-based algorithms use $f^j_i$ values to produce an estimation. The computational resources required to obtain $f^j_i$ from sampled packets are far greater than those required for calculating the estimation itself. Hence, to measure how much CPU is consumed by our framework, and to compare it to statistical sampling-based algorithms, we simply count the number of processed packets, while ignoring the CPU consumption for the execution of \emph{predict()} and \emph{partial\_fit()}. In the same way, we ignore all calculations performed by the statistical sampling-based algorithms to which we compare our framework. 
		
		Statistical sampling-based algorithms process only the sampled packets, while our framework processes all sampled packets as well as complete training batches. Thus, in a statistical sampling-based algorithm the expected number of  processed packets in a batch is 
		\begin{align*}
		\emph{mean\_batch\_size} \cdot \emph{sampling\_rate},
		\end{align*}
		while in our framework, this number is
		\begin{align*}
		\emph{mean\_batch\_size} \cdot \emph{effective\_sampling\_rate}
		\end{align*}
		where,
		\begin{align*}
		\emph{effective\_sampling\_rate}=& \emph{sampling\_rate}+\emph{training\_rate}\\&
		-\emph{sampling\_rate} \cdot \emph{training\_rate}.
		\end{align*}
		Thus, the computational cost of statistical sampling-based algorithms is proportional to the \emph{sampling\_rate}, and the computational cost of our framework is proportional to the \emph{effective\_sampling\_rate}. When we compare between the two methods in Section \ref{sec:evaluation}, we ensure that the \emph{sampling\_rate} for the sampling-based algorithms is equal to the \emph{effective\_sampling\_rate} of our framework.
	
	\subsection{Implementation Details}

			Our framework is designed to perform real-time cardinality estimation. However, in order to get reproducible and comparable results over all estimation methods, in our experiments data processing was performed in advance as follows\footnote{The code used for our experiments can be found at \url{https://github.com/yuvalnezri/CardEst}.}. First, all batches were sampled. Then, features and statistical properties were extracted from these samples, and the true cardinality of each batch was calculated. Then, these values were used to calculate an estimation and to train the online ML models. This guarantees that all estimators and online ML algorithms use the exact same samples and features for their estimations.
			
			In steps \ref{op:train_phase}--\ref{op:end_phase} of Procedure \ref{alg:model}, batch's true cardinality is computed concurrently with the \emph{partial\_fit()} operation. This concurrent execution prevents the training phase from delaying subsequent estimations. In our experiments this delay was insignificant, since the true cardinality of each batch was calculated in advance. Nonetheless, to express this behavior, during each training phase we first provided an estimation and only then trained the online ML model. Thus, it is assumed that the training phase ends before the subsequent batch estimation is requested.
			
			Online ML algorithms rely on training over previously seen data examples, which are usually unavailable during the first stages of the model's lifespan. Hence, some kind of initialization process is required. This process of initializing the online ML algorithms is usually referred to as ``bootstrapping''. To bootstrap an online ML model we always used the first batch of packets as a training batch. Since each algorithm was implemented differently, a different bootstrapping process was used for each. For SGD we ran numerous \emph{partial\_fit()} iterations until the loss difference between two subsequent iterations became smaller than a predefined tolerance value, while for PA and RLS we found that a single \emph{partial\_fit()} operation is sufficient.

	\subsection{The CAIDA-2016 Trace} \label{subsec:caida-2016}
		
		The CAIDA-2016 \cite{caidabackbone2016} trace was collected from the Equinix-Chicago high-speed monitor over Internet backbone links. The part of this trace that we use contains 44,567,284 packets, collected during approximately 48 seconds.
		
		Our analysis of the frequency distribution ($f_i^j$ values) of the first batch shows a heavy-tailed behavior: 81\% of the flows are small (contain less than 4 packets), while 51\% of the packets belong to big flows (contain more than 20 packets). Other batches show similar distribution. Statistical sampling-based algorithms are known to demonstrate high error rates in heavy-tail distributed traffic \cite{charikar2000towards, haas1998estimating}.
		
\begin{figure}[!tb]
			\centering
			\includegraphics[width=.99\linewidth]{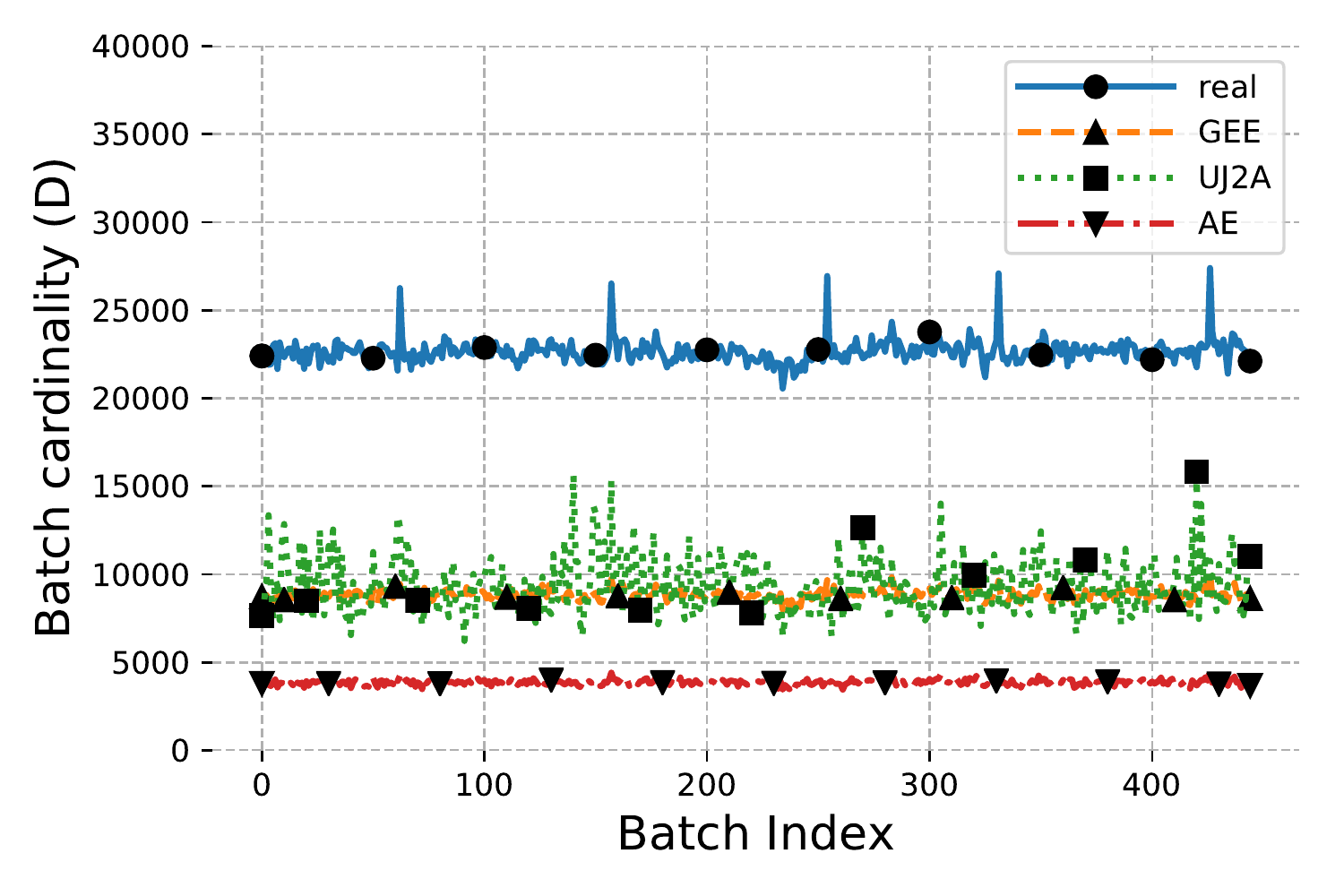}
			\caption{Real vs. estimated batch cardinality of statistical sampling-based algorithms for the CAIDA-2016 trace}
			\label{fig:caida-2016_statistical}
		\end{figure}

		Figure \ref{fig:caida-2016_statistical} shows real vs. estimated cardinality, when the estimation is performed using 3 statistical sampling-based algorithms: GEE, AE and UJ2A. For these graphs, the \emph{batch\_size} is 100K and the \emph{sampling\_rate} is $q=0.0199$. We can see that the prediction of all algorithms is far from the real cardinality.

\begin{figure}[!tb]
			\centering
			\includegraphics[width=.99\linewidth]{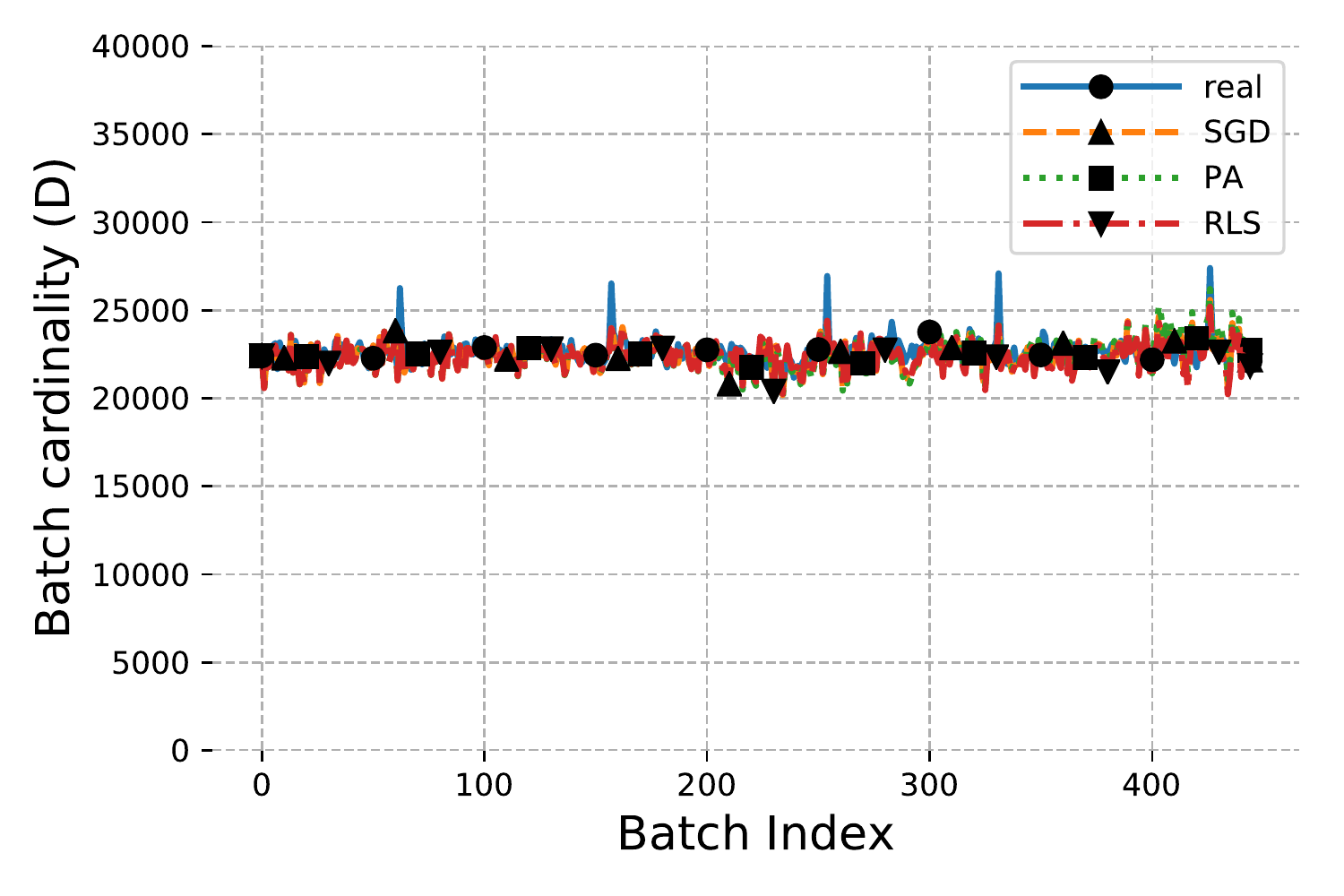}
			\caption{Real vs. estimated batch cardinality of the proposed framework with different online ML algorithms for the CAIDA-2016 trace}
			\label{fig:caida-2016_online_ml}
		\end{figure}

		Figure \ref{fig:caida-2016_online_ml} shows real vs. estimated cardinality when the estimation is performed using our framework with 3 different online linear regression ML algorithms: Stochastic Gradient Descent (SGD), Recursive Least Squares (RLS) and Passive Aggressive (PA). In this graph we use $\emph{batch\_size}=100K$, $\emph{q}=0.01$ and $\emph{training\_rate}=0.01$. The feature set is only $\{f_i^1\}$. These parameters yield an \emph{effective\_sampling\_rate} of 0.0199, which is identical to that used in Figure \ref{fig:caida-2016_statistical} for the sampling-based algorithms. In addition, we set the \emph{learning\_rate} of SGD to $10^{-6}$, the forgetting factor of RLS to $\mu=0.99$, the $\epsilon$-insensitive loss function of PA to $\epsilon=0.1$, and the aggressiveness parameter of PA-II update rule to $C=1$. It can be clearly seen that the prediction of all the algorithms used in our framework is very close to the real cardinality.
		
		Figure \ref{fig:caida-2016_error} summarizes the error rates of the experiments reported in Figure \ref{fig:caida-2016_statistical} and \ref{fig:caida-2016_online_ml}. Indeed, we see that the error rates for the statistical sampling-based algorithms are significantly higher than those of our new framework, regardless of the online ML algorithms being used. This is despite the fact that the same number of packets are processed in all cases. In addition, we can see that when the flow size distribution hardly changes, the various online ML algorithms show similar accuracy. We also see that for this specific trace, AE is significantly less accurate than GEE and UJ2A. Overall, the proposed framework's MAPE is approximately 30 times better than the best performing statistical sampling-based algorithm.

		\begin{figure}
			\centering
			\small
			\begin{tabular}{ccccc}
\toprule
{} &      RMSE &       MAE &  MAPE &     MAXAE \\
\midrule
GEE  &  13,803.0 &  13,793.6 &  60.9 &  17,441.7 \\
AE   &  18,789.0 &  18,780.2 &  82.9 &  22,878.7 \\
UJ2A &  13,449.6 &  13,347.0 &  58.9 &  18,629.6 \\
SGD  &     703.7 &     550.7 &   2.4 &   2,872.7 \\
PA   &     751.3 &     585.0 &   2.6 &   2,798.0 \\
RLS  &     704.0 &     549.2 &   2.4 &   2,956.7 \\
\bottomrule
\end{tabular}

			\caption{Different error metrics of the various sampling-based and ML estimators for the CAIDA-2016 trace. The algorithms used in our framework perform significantly better}
			\label{fig:caida-2016_error}
		\end{figure}
		
\begin{figure}[!tb]
			\centering
			\includegraphics[width=.49\textwidth]{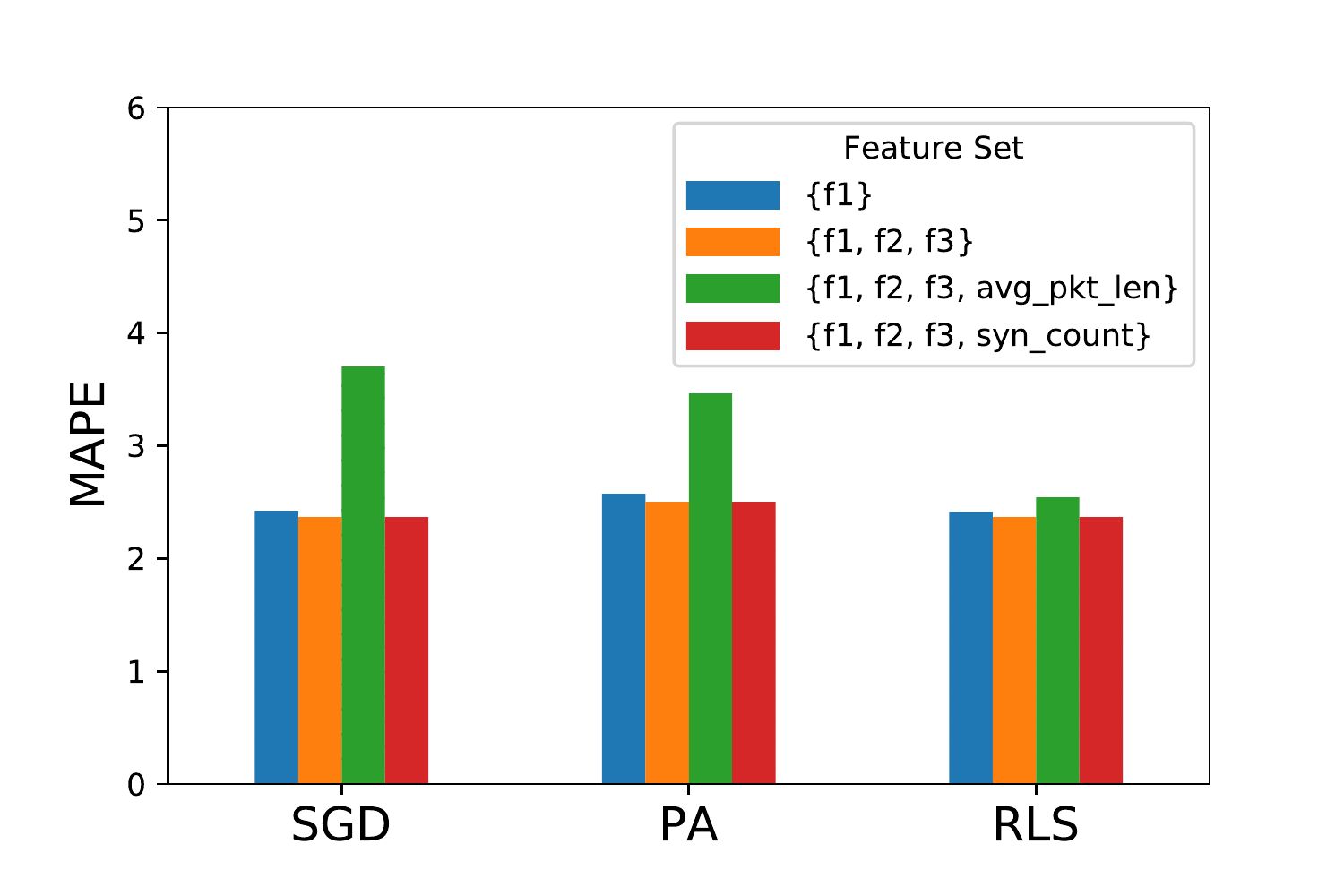}
			\caption{The impact of extending the feature set on the accuracy of our framework for the CAIDA-2016 trace}
			\label{fig:caida-2016_features}
		\end{figure}

		Next, we examine the effect of adding more features to the online ML algorithm, as discussed in Section \ref{subsec: feature_select}. Figure \ref{fig:caida-2016_features} depicts the MAPE of the three online ML algorithms with the following feature sets: $\{f_i^1\}$, $\{f_i^1, f_i^2, f_i^3\}$,  $\{f_i^1, f_i^2, f_i^3, avg\_pkt\_len\}$ and $\{f_i^1, f_i^2, f_i^3, syn\_count\}$. We conclude from this graph that extending the features beyond $\{f_i^1\}$ does not yield significant improvement in the accuracy. Moreover, adding \emph{avg\_pkt\_len} to the feature set even slightly increases the error. 
		
\begin{figure}[!t]
			\centering
			\includegraphics[width=.49\textwidth]{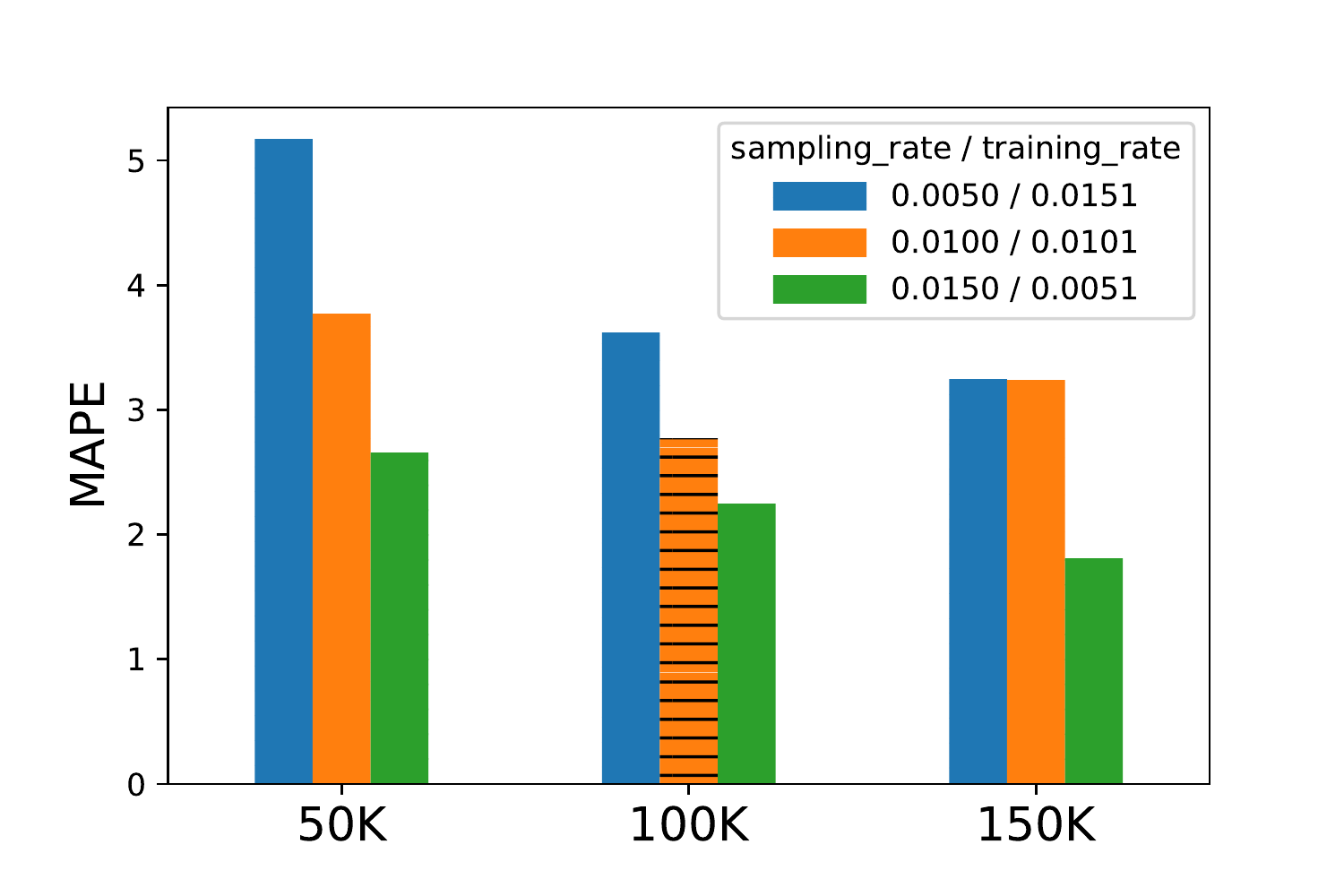}
			\caption{The trade-off between \emph{sampling\_rate} and \emph{training\_rate} for the CAIDA-2016 trace, with fixed \emph{effective\_sampling\_rate} of 2\%} 
			\label{fig:caida-2016_sampling_training}
		\end{figure}

		As described in Section \ref{subsec:effective}, \emph{effective\_sampling\_rate} is our metric for quantifying the computational cost of our framework. To examine the trade-off between \emph{sampling\_rate} and \emph{training\_rate}, we fix \emph{effective\_sampling\_rate} to 0.02 (i.e.\, 2\% of the stream packets are processed), and vary the \emph{sampling\_rate} between 0.005 and 0.015. We then set \emph{training\_rate} such that the \emph{effective\_sampling\_rate} will be 0.02. Figure \ref{fig:caida-2016_sampling_training} shows the MAPE of this experiment with 3 batch sizes when PA is used as the online ML algorithm. 
		
		It is evident from Figure \ref{fig:caida-2016_sampling_training} that bigger batches yield better accuracy than smaller batches: as the batch size increases, the sample becomes more statistically representative, and the correlation between $f_i^1$ and the real batch cardinality increases. 
		
		We can also see that it is better to use a high sampling rate with a low training rate rather than vice versa. This is because the cardinality of this trace only changes slightly over time. Hence, more frequent training does not improve the accuracy over time, and it is better to invest packet processing resources in sampling more packets in every batch rather than in more training batches.
		
	\subsection{The CAIDA-DDoS Trace}

		CAIDA-DDoS trace \cite{caidaddos2007} contains approximately one hour of anonymized traffic from a Distributed Denial-of-Service (DDoS) attack that occurred on August 4th, 2007. We use 26,760,675 packets from this trace. These packets represent approximately 300 seconds, and they include the transition from ``no-attack'' to ``attack''. Since DDoS detection is a time sensitive application, our \emph{estimation\_rate} is a function of time rather than number of packets.
		
\begin{figure}[!tb]
			\centering
			\includegraphics[width=.95\linewidth]{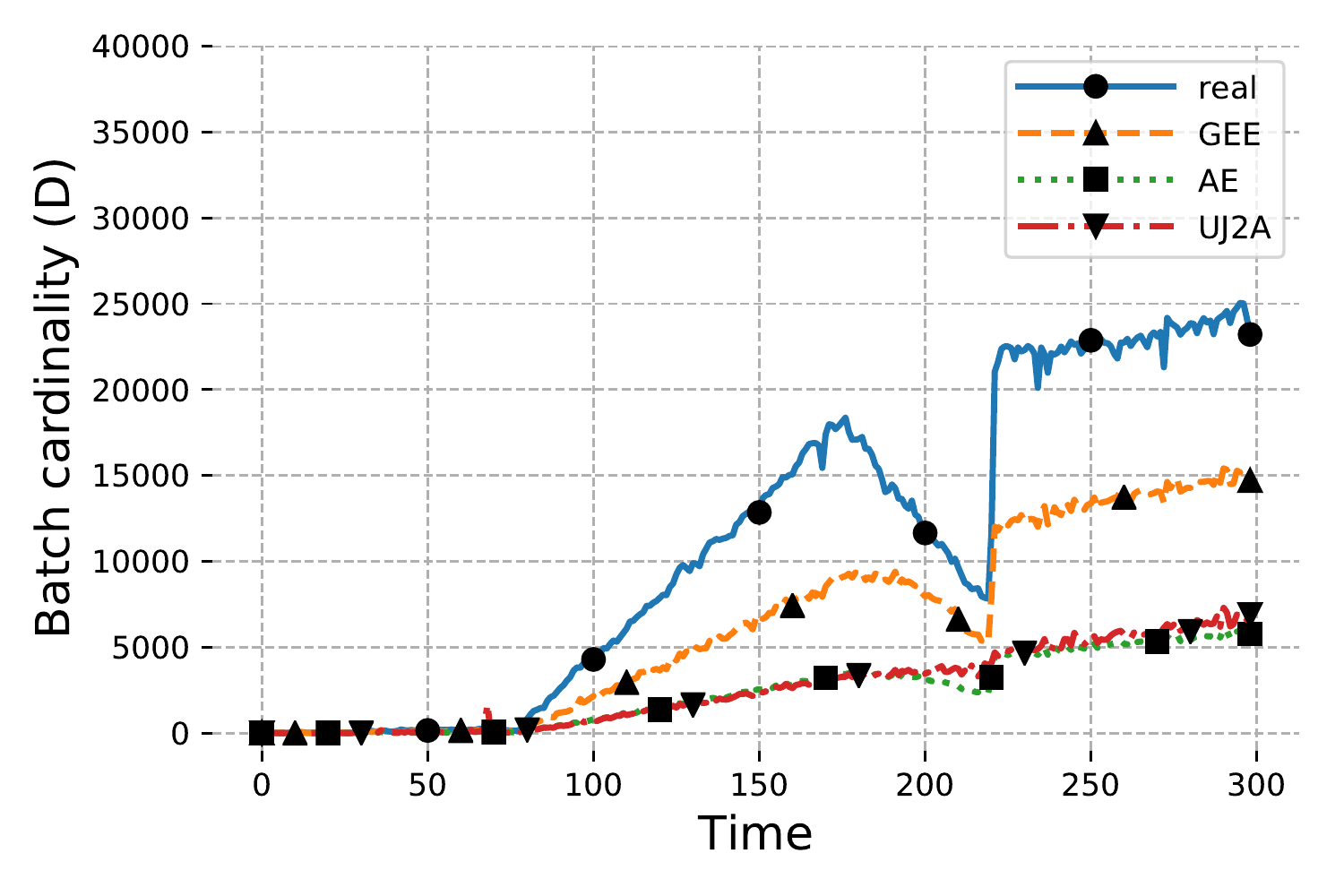}
			\caption{Real vs. estimated batch cardinality of statistical sampling-based algorithms for the CAIDA-DDoS trace}
			\label{fig:caida-ddos_statistical}
		\end{figure}

		After the attack begins, the flow size distribution is heavy-tailed. Thus, as for the CAIDA-2016 trace, we expect a relatively high error rate when statistical sampling-based algorithms are used. Figure \ref{fig:caida-ddos_statistical} shows the real cardinality vs. the cardinality estimated by statistical sampling-based algorithms, with \emph{batch\_size} of 1 second and sampling rate of $q=0.0199$.

		\begin{figure}[!tb]
			\centering
			\includegraphics[width=.95\linewidth]{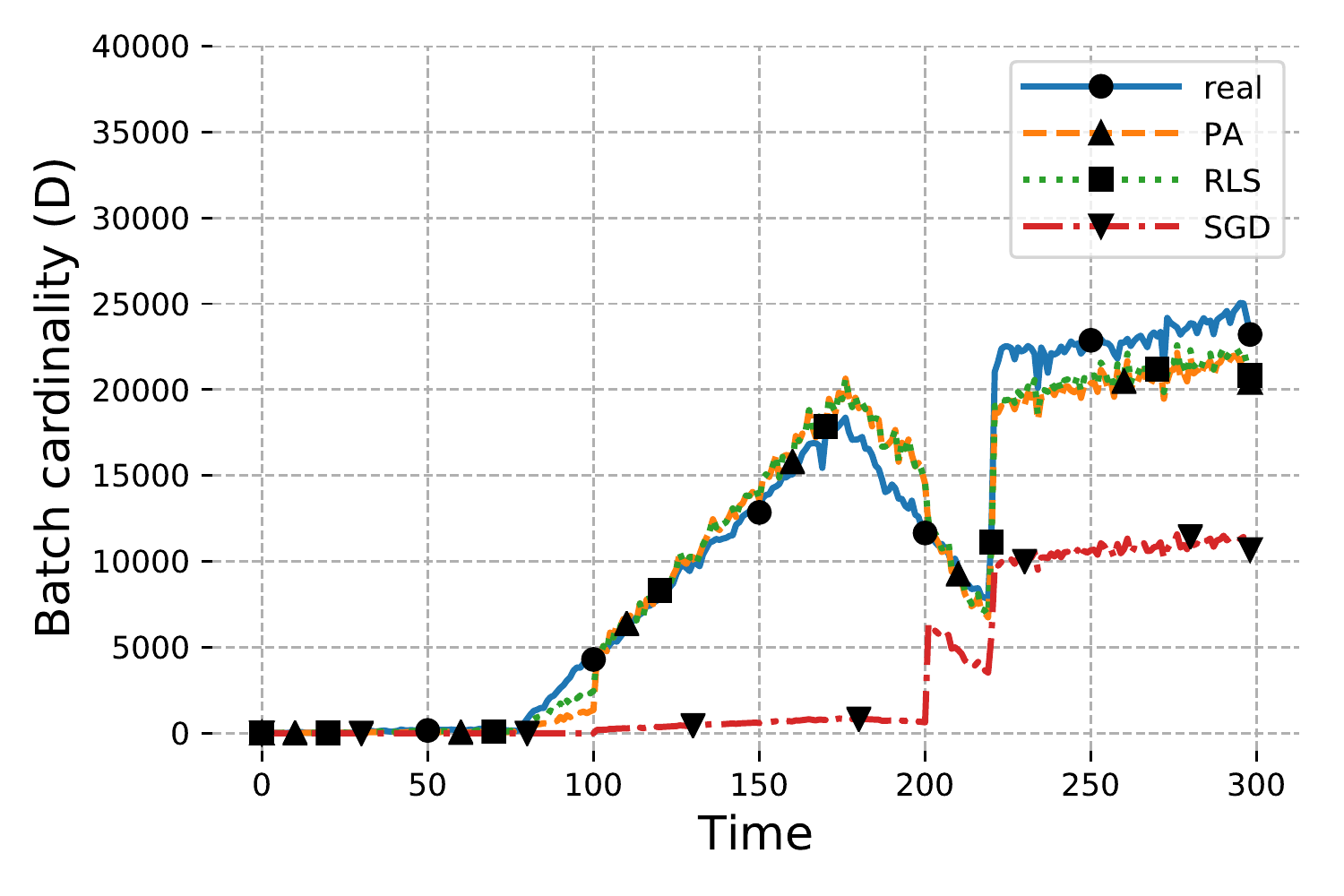}
			\caption{Real vs. estimated batch cardinality of the proposed framework with different online ML algorithms for the CAIDA-DDoS trace}
			\label{fig:caida-ddos_online_ml}
		\end{figure}

		\begin{figure}
			\centering
			\small
			\begin{tabular}{ccccc}
\toprule
{} &      RMSE &      MAE &  MAPE &     MAXAE \\
\midrule
GEE  &   5,976.9 &  4,667.7 &  41.8 &  10,422.8 \\
AE   &  11,004.0 &  8,539.6 &  75.1 &  19,141.3 \\
UJ2A &  10,774.4 &  8,362.3 &  73.8 &  18,240.5 \\
SGD  &   9,548.2 &  7,623.9 &  82.9 &  17,474.7 \\
PA   &   1,689.7 &  1,260.7 &  25.4 &   4,024.1 \\
RLS  &   1,485.2 &  1,100.2 &  21.4 &   4,043.9 \\
\bottomrule
\end{tabular}

			\caption{Different error metrics of the various sampling-based and ML estimators for the CAIDA-DDoS trace}
			\label{fig:caida-ddos_error}
		\end{figure}
		
		Next, we use CAIDA-DDoS to test our framework with different online ML algorithms and \emph{batch\_size} of 1 second. We set $\emph{sampling\_rate} = 0.01$, $\emph{training\_rate} = 0.01$, and use only $\{f_i^1\}$ as our feature set. These parameters yield an effective sampling rate of 0.0199, which is the same as that used for the sampling-based algorithms in Figure \ref{fig:caida-ddos_statistical}. The ML-specific parameters are similar to those used for Figure \ref{fig:caida-2016_online_ml}: SGD has a \emph{learning\_rate} of $10^{-6}$, RLS has a forgetting factor of $\mu=0.99$, PA has an $\epsilon$-insensitive loss function of $\epsilon=0.1$, and the PA-II update rule has an aggressiveness parameter of $C=1$. Figure \ref{fig:caida-ddos_online_ml} describes the real and estimated cardinality for each online ML algorithm. 
		
		Figure \ref{fig:caida-ddos_error} summarizes the error rates of Figure \ref{fig:caida-ddos_statistical} vs. \ref{fig:caida-ddos_online_ml}. We can conclude that SGD's inherent \emph{learning\_rate} is too low. Thus, it fails to adapt fast enough to changes in flow size distribution. But when we increase its \emph{learning\_rate} to $10^{-5}$, the estimation diverges after the attack starts and shows even higher error rates. SGD is known for its \emph{learning\_rate} sensitivity, and it is therefore not recommended when cardinality values are expected to change drastically. In contrast to SGD, PA and RLS are significantly better than the statistical sampling-based algorithms and we see 50\% improvement in the MAPE of the best online ML algorithm (RLS), over the best statistical sampling-based algorithm. As for the first trace, also in this case we do not see significant impact by extending the feature set beyond ${f^1_i}$.
		
\begin{figure}[!t]
			\centering
			\includegraphics[width=.49\textwidth]{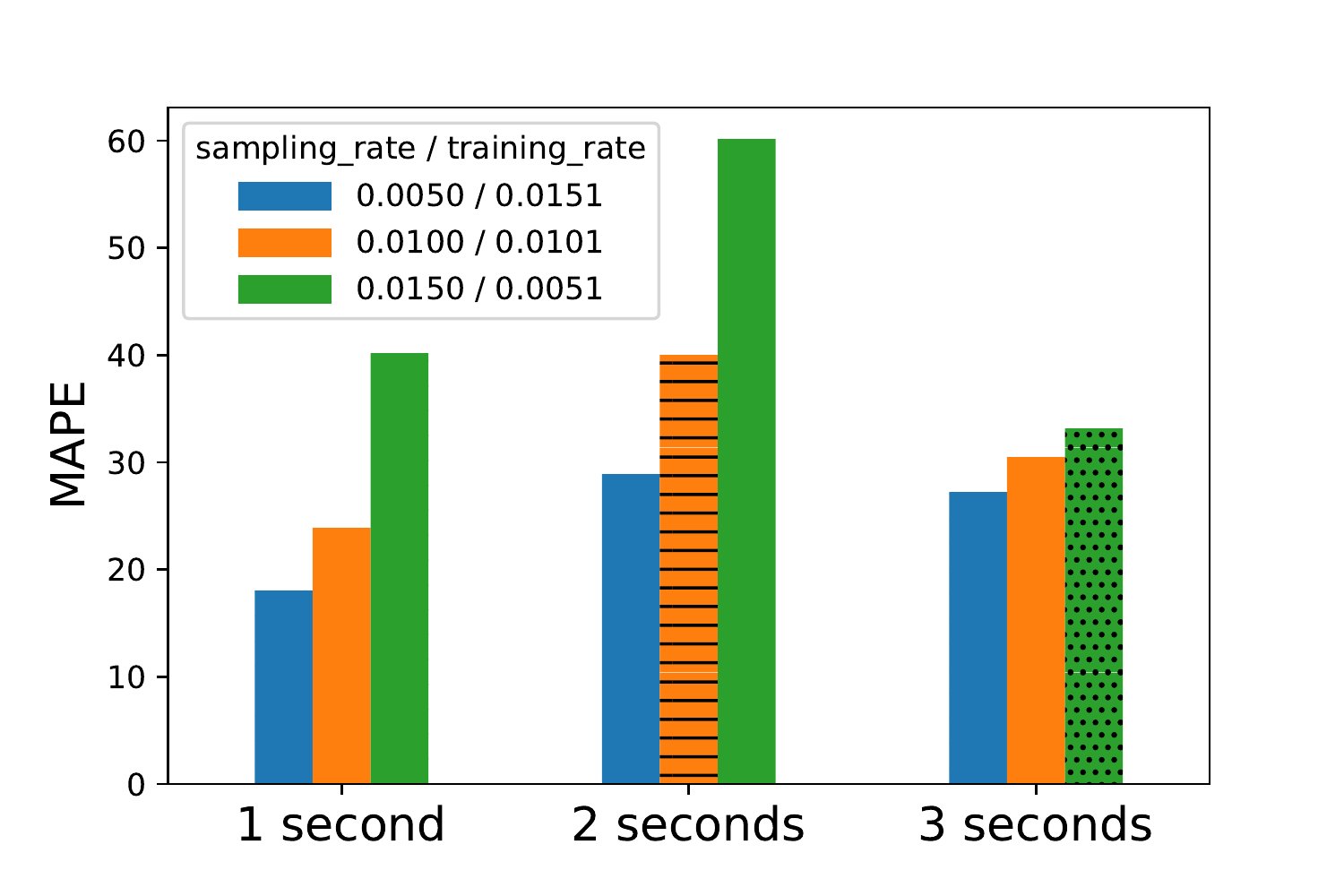}
			\caption{The trade-off between \emph{sampling\_rate} and \emph{training\_rate} for the CAIDA-DDoS trace, with fixed \emph{effective\_sampling\_rate} of 2\%}
			\label{fig:caida-ddos_sampling_training}
		\end{figure}

		Next, we examine the trade-off between \emph{sampling\_rate} and \emph{training\_rate}. We set a fixed \emph{effective\_sampling\_rate} of 0.02 (2\%), and use sampling rates of 0.005 (0.5\%), 0.01 (1\%) and 0.015 (1.5\%). The training rate is calculated such that the \emph{effective\_sampling\_rate} remains 0.02. Figure \ref{fig:caida-ddos_sampling_training} shows the MAPE for the various algorithms for three batch sizes. It is evident that increasing the sampling rate while decreasing the training rate yields better accuracy. This is in contrast to what we saw in Figure \ref{fig:caida-2016_sampling_training} for the CAIDA-2016 trace. In the CAIDA-DDoS trace, the flow size distribution changes frequently. Hence, a higher \emph{training\_rate} is required to obtain high accuracy.

	\subsection{The DARPA-DDoS Trace}

		The DARPA-DDoS trace \cite{darpaddos} contains a SYN flood DDoS attack on one target (IP address 172.28.4.7) and some background traffic. The DDoS traffic comes from about 100 different sources, some of which contribute significantly more traffic than the others.
		
		We use a part of the trace in which the attack occasionally goes on and off. It contains 3,955,270 packets, collected during 352 seconds. Since the number of packets is relatively small, a sampling rate of 1\% or 2\% yields a not statistically representative \emph{sample\_size}, and very high error rates. Hence, for this trace we use higher sampling rates.%
		
\begin{figure}[!tb]
			\centering
			\includegraphics[width=.95\linewidth]{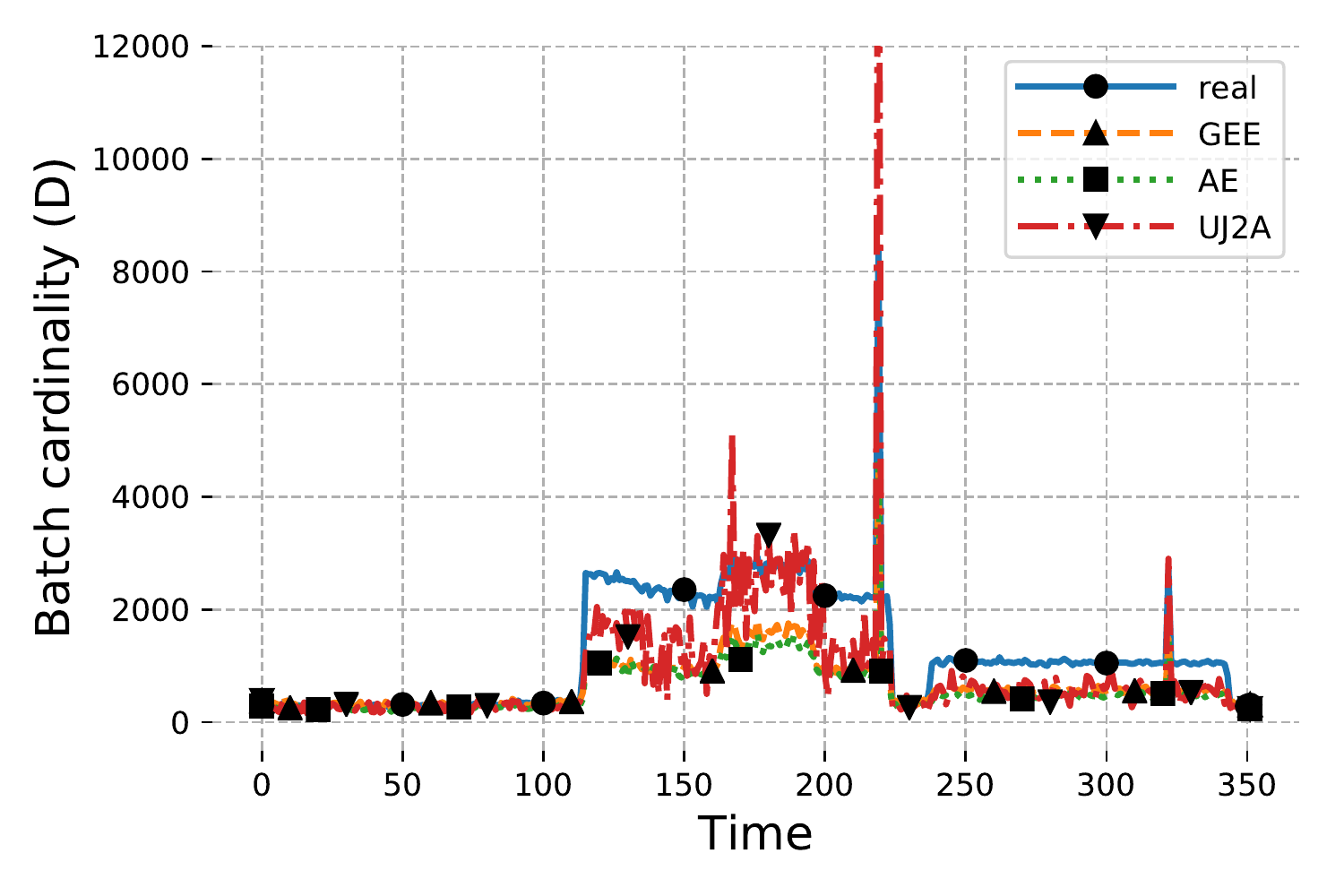}
			\caption{Real vs. estimated batch cardinality of statistical sampling-based algorithms for the DARPA-DDoS trace}
			\label{fig:darpa-ddos_statistical}
		\end{figure}

\begin{figure}[!tb]
			\centering
			\includegraphics[width=.95\linewidth]{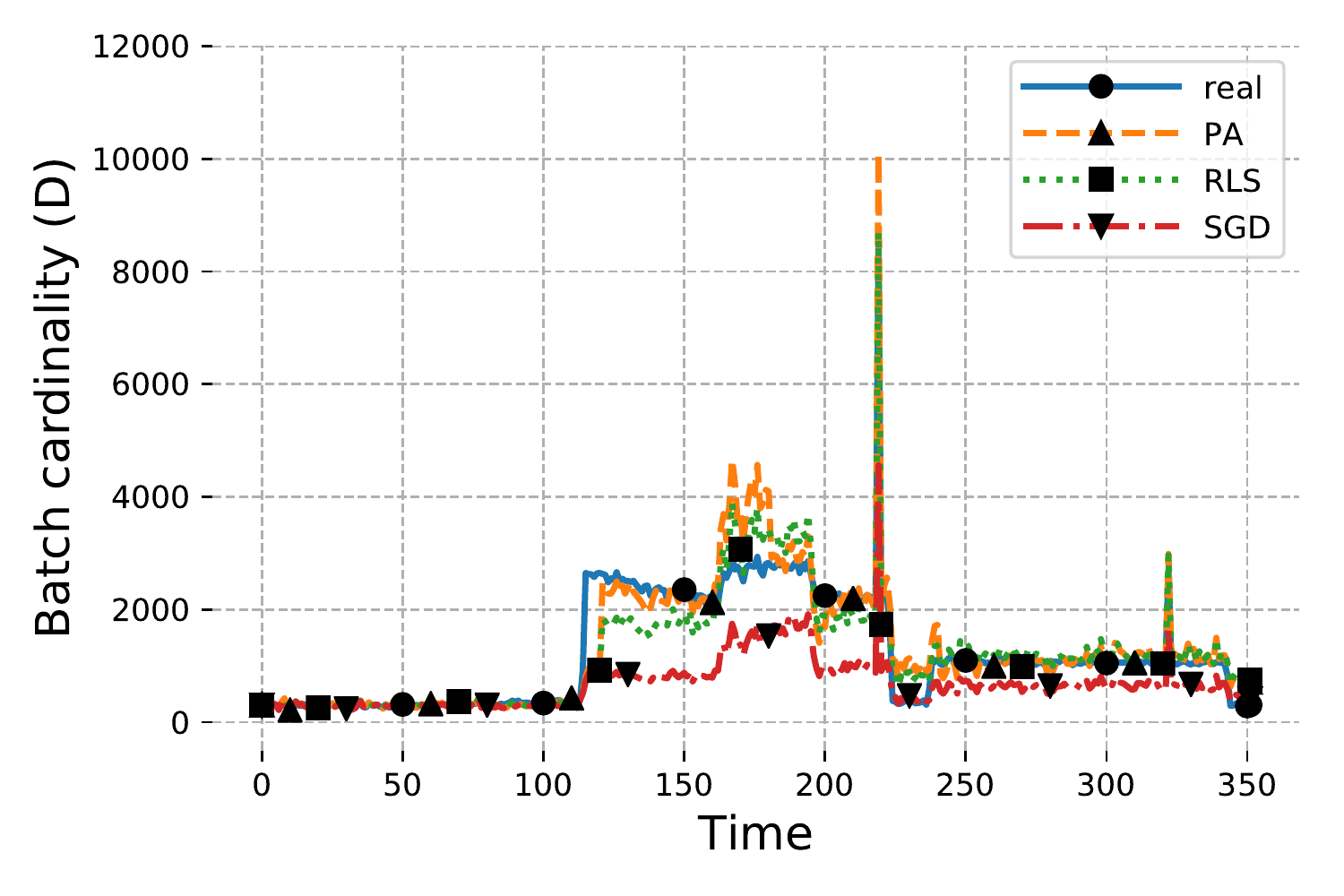}
			\caption{Real vs. estimated batch cardinality of the proposed framework with different online ML algorithms for the DARPA-DDoS trace}
			\label{fig:darpa-ddos_online_ml}
		\end{figure}

		Figure \ref{fig:darpa-ddos_statistical} shows the real vs. estimated cardinality of 3 statistical sampling-based algorithms with \emph{batch\_size} of 1 second and \emph{sampling\_rate} of $q=0.975$. Figure \ref{fig:darpa-ddos_online_ml} shows the real vs. estimated cardinality of the proposed framework with the three online ML algorithms. The framework parameters used in Figure \ref{fig:darpa-ddos_online_ml} are: $\emph{batch\_size} = 1$ second, $\emph{sampling\_rate} = 0.05$ and $\emph{training\_rate} = 0.05$. These parameters yield an \emph{effective\_sampling\_rate} of 0.975, which is the same as that used for the statistical sampling-based algorithms in Figure \ref{fig:darpa-ddos_statistical}. The feature set consists of $\{f_i^1\}$ only. As before, SGD's \emph{learning\_rate} is set to $10^{-6}$, RLS's forgetting factor is set to $\mu=0.99$, PA's $\epsilon$-insensitive loss function is set to $\epsilon=0.1$, and the PA-II update rule aggressiveness parameter is set to $C=1$.
		
		\begin{figure}
			\centering
			\small
			\begin{tabular}{ccccc}
\toprule
{} &   RMSE &    MAE &  MAPE &     MAXAE \\
\midrule
GEE  &  800.9 &  569.6 &  33.8 &   3,898.1 \\
AE   &  873.4 &  645.0 &  41.4 &   3,729.8 \\
UJ2A &  829.4 &  452.0 &  31.2 &  10,826.3 \\
SGD  &  447.4 &  271.1 &  23.1 &   1,907.2 \\
PA   &  420.0 &  220.0 &  22.1 &   1,844.3 \\
RLS  &  404.0 &  259.2 &  22.4 &   1,884.8 \\
\bottomrule
\end{tabular}

			\caption{Different error metrics of the various sampling-based and ML estimators for the DARPA-DDoS trace}
			\label{fig:darpa-ddos_error}
		\end{figure}
		
		Figure \ref{fig:darpa-ddos_error} summarizes the error rates of the above experiments. We can see that UJ2A's MAXAE rate is much higher than that of the other statistical sampling-based algorithms. This error is obtained from batch no. 219, which presents an extremely high cardinality value. Overall we see an improvement of ~30\% in the MAPE of the best online ML algorithm (PA) over the best statistical sampling-based algorithm (UJ2A).
		
\begin{figure}[!tb]
			\centering
			\includegraphics[width=.49\textwidth]{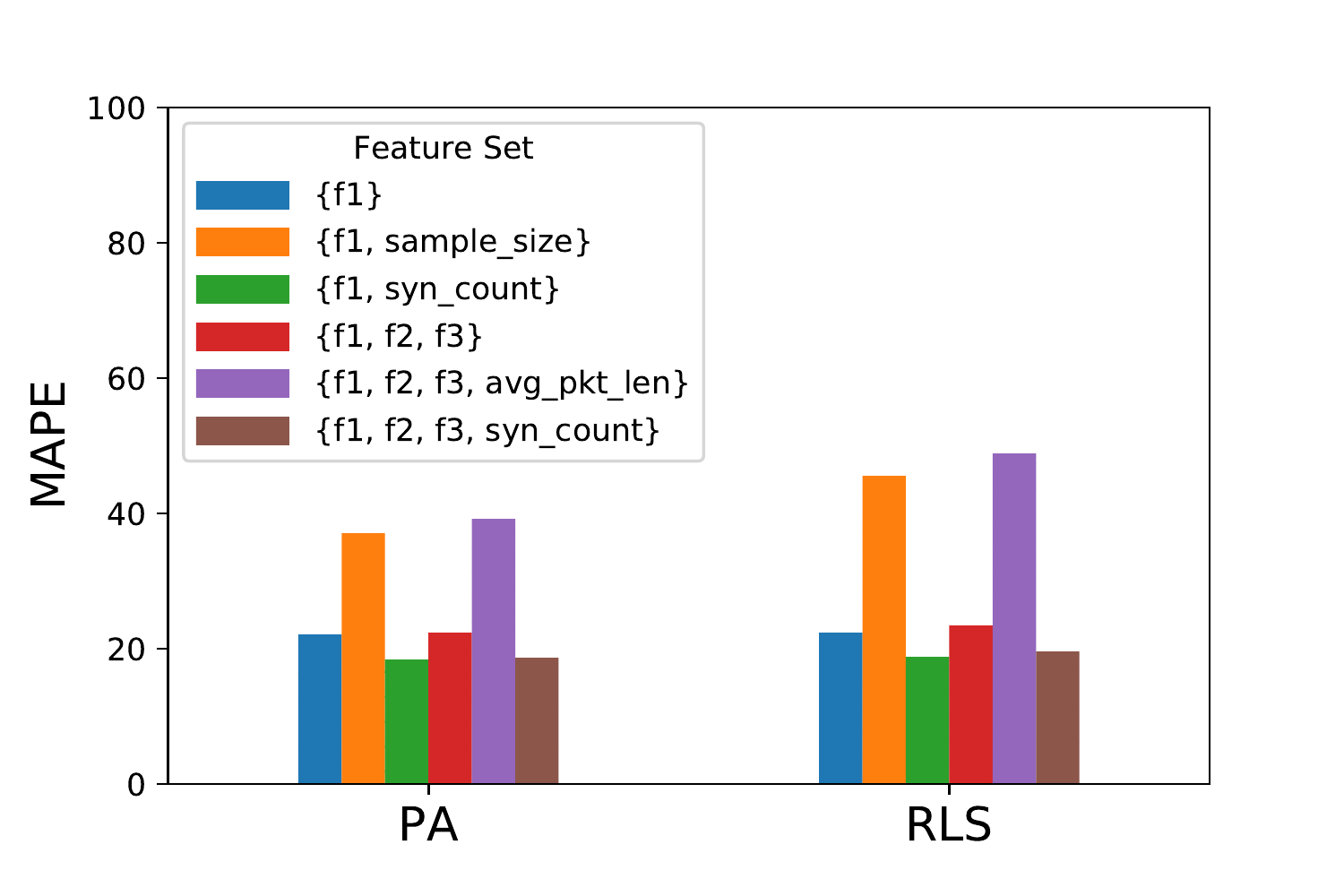}
			\caption{The impact of extending the feature set on the accuracy of our framework for the DARPA-DDoS trace}
			\label{fig:darpa-ddos_features}
		\end{figure}

		Figure \ref{fig:darpa-ddos_features} presents the accuracy of PA and RLS with different feature sets. Since this trace contains a SYN attack, feature set $\{f_i^1, \emph{syn\_count}\}$ slightly improves the MAPE.
		
	\subsection{The UCLA-CSD Trace}

		The UCLA-CSD trace \cite{uclacsd} contains packets collected during August 2001 at the border router of the UCLA Computer Science Department. The part of this trace that we use contains 30,000,000 TCP packets, collected during approximately 14 hours.
			
		\begin{figure}[!tb]
			\centering
			\includegraphics[width=.95\linewidth]{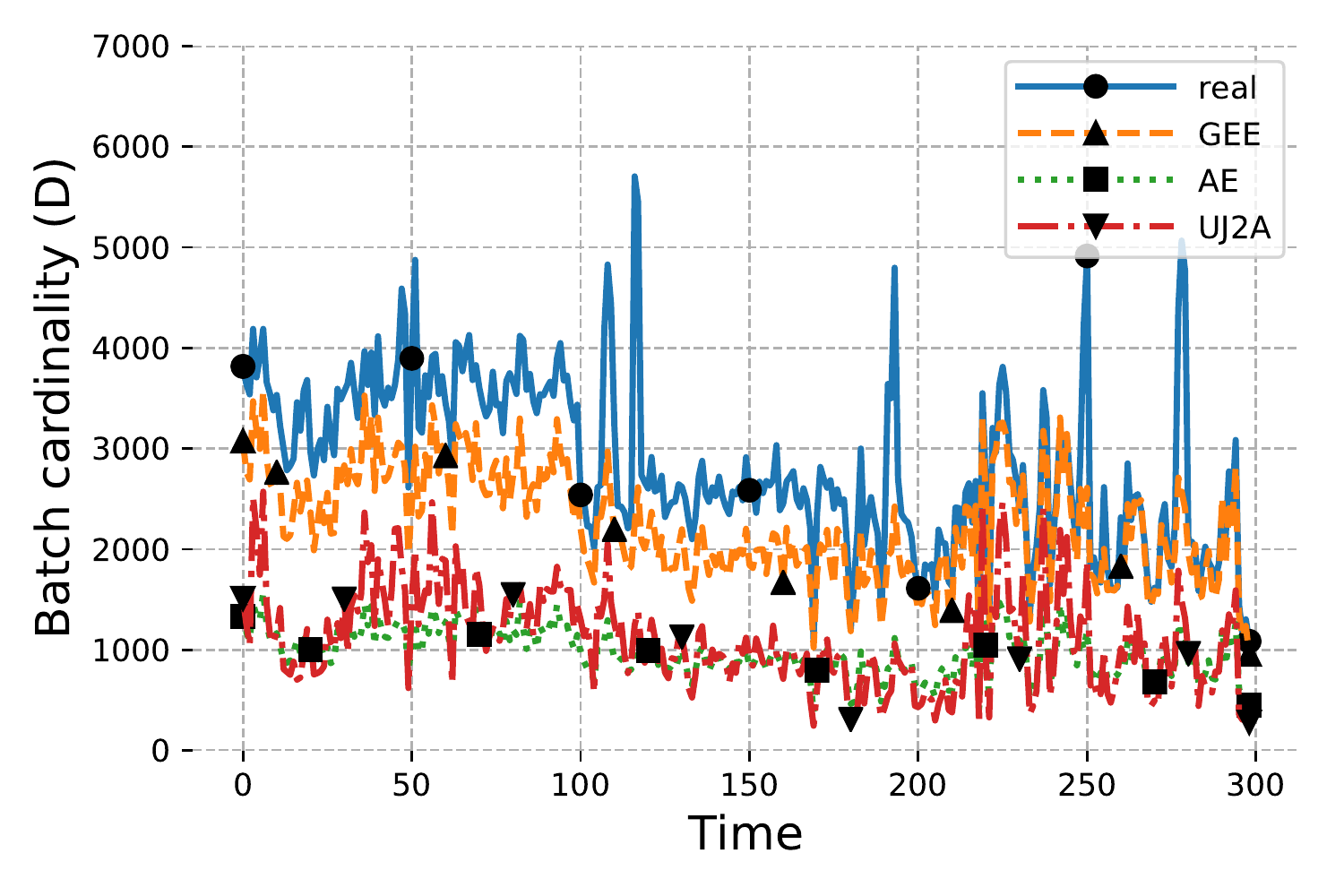}
			\caption{Real vs. estimated batch cardinality of statistical sampling-based algorithms for the UCLA-CSD trace}
			\label{fig:ucla-csd_statistical}
		\end{figure}
		
		\begin{figure}[!tb]
			\centering
			\includegraphics[width=.95\linewidth]{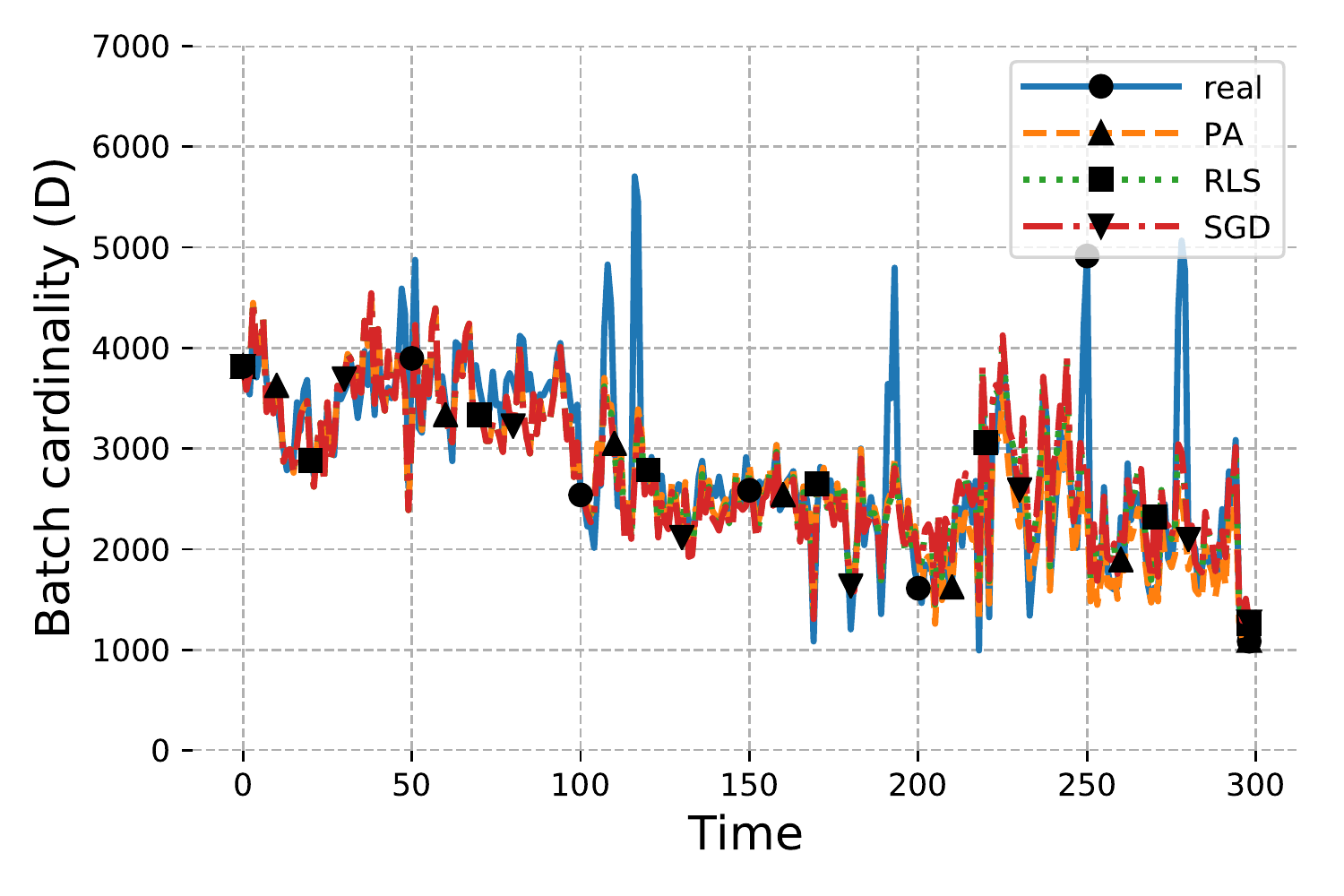}
			\caption{Real vs. estimated batch cardinality of the proposed framework with different online ML algorithms for the UCLA-CSD trace}
			\label{fig:ucla-csd_online_ml}
		\end{figure}

		Figure \ref{fig:ucla-csd_statistical} shows the real vs. estimated cardinality of the three statistical sampling-based algorithms with \emph{batch\_size} of 100K packets and \emph{sampling\_rate} of $q=0.0199$. UCLA-CSD also shows a heavy-tailed behavior, but its skewness is less acute than that of the CAIDA-2016 trace (Section \ref{subsec:caida-2016}). For example, analysis of the first batch shows that only 37\% of the packets belong to small flows (that contain less than 4 packets). This fact can explain the performance advantage of GEE compared to AE and UJ2A.		
		
		\begin{figure}
			\centering
			\small
			\begin{tabular}{ccccc}
\toprule
{} &     RMSE &      MAE &  MAPE &    MAXAE \\
\midrule
GEE  &    763.7 &    600.2 &  19.2 &  3,458.1 \\
AE   &  1,977.4 &  1,863.6 &  64.7 &  4,720.7 \\
UJ2A &  1,824.0 &  1,735.7 &  61.9 &  4,460.2 \\
SGD  &    457.0 &    277.2 &   9.7 &  2,972.0 \\
PA   &    474.0 &    271.8 &   9.0 &  2,910.8 \\
RLS  &    457.9 &    279.4 &   9.8 &  2,974.2 \\
\bottomrule
\end{tabular}

			\caption{Different error metrics of the various sampling-based and ML estimators for the UCLA-CSD trace}
			\label{fig:ucla-csd_error}
		\end{figure}
		
		Figure \ref{fig:ucla-csd_online_ml} shows the true cardinality vs. estimated cardinality values for our framework with each online ML algorithm. For this comparison we set the following framework parameters: $\emph{batch\_size} = \text{100K}$, $q=0.01$, $\emph{training\_rate} = 0.01$. The feature set contains only $\{f_i^1\}$. These parameters yield an effective sampling rate of 0.0199, which is identical to what we used in Figure \ref{fig:ucla-csd_statistical} for the sampling-based algorithms. SGD's \emph{learning\_rate} is set to $10^{-5}$, since it shows the smallest error. RLS's forgetting factor is $\mu=0.99$, PA uses the $\epsilon$-insensitive loss function with $\epsilon=0.1$, and the PA-II update rule with an aggressiveness parameter of $C=1$.
		
\begin{figure}[!tb]
			\centering
			\includegraphics[width=.49\textwidth]{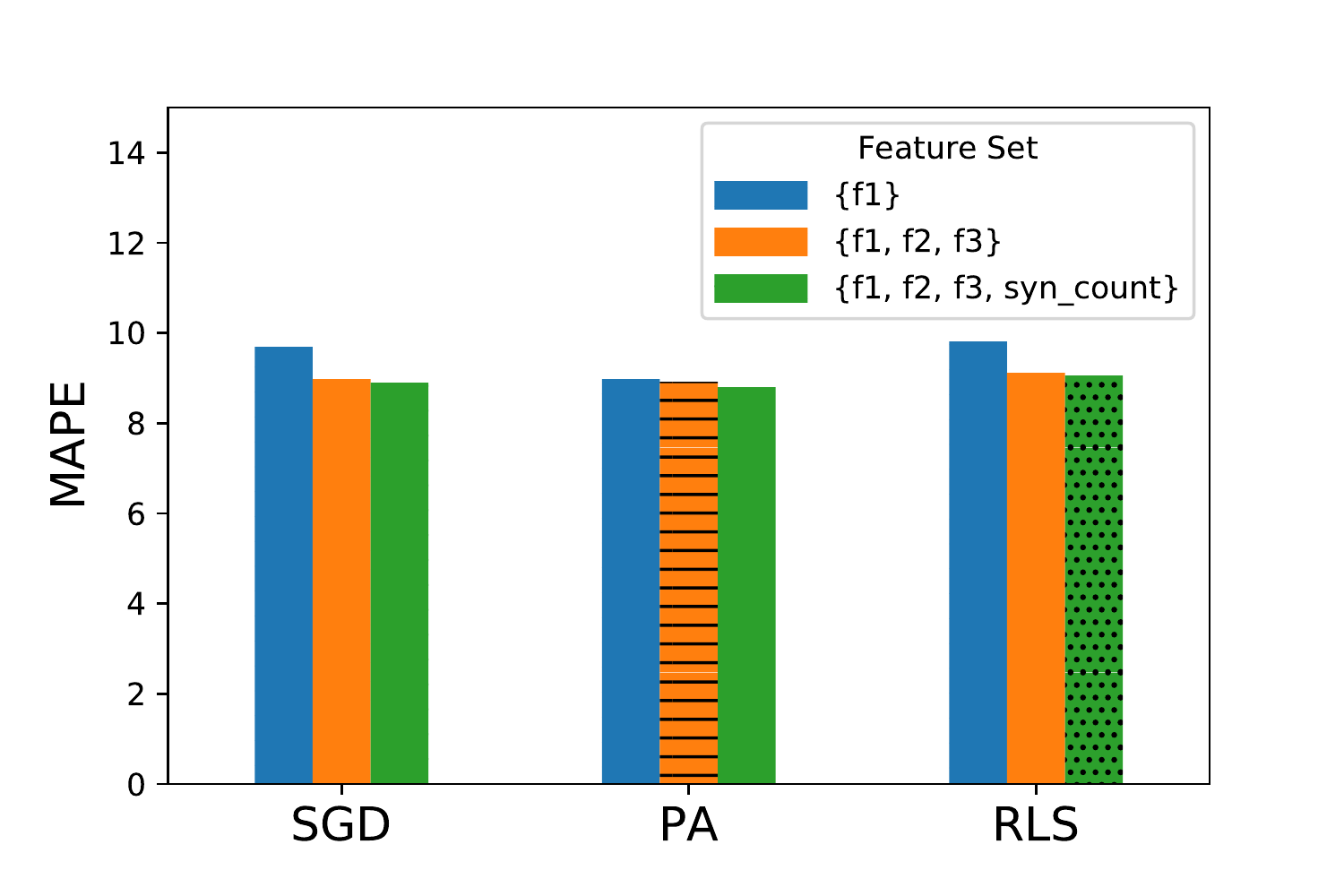}
			\caption{The impact of extending the feature set on the accuracy of our framework for the UCLA-CSD trace}
			\label{fig:ucla-csd_features}
		\end{figure}

		Figure \ref{fig:ucla-csd_error} summarizes the error rates of the above experiments. Our framework obtains an MAPE improvement of ~10\% over the best statistical sampling-based algorithm. Figure \ref{fig:ucla-csd_features} presents the accuracy of PA and RLS with different feature sets.
		
	\subsection{Computational Cost Analysis} \label{sec:computational_cost}
	
		Although calculating the estimation itself is not the most resource intensive operation, we present the following evaluation of its computational cost. First, we analytically analyze the runtime complexity of the computations carried out by the sampling-based algorithms, and that of the \emph{predict()} and \emph{partial\_fit()} operations carried out by our framework. Then, we show empirical measurements to support this analysis.
		
		Except for AE, which uses a numerical method in its estimation process and is more computationally expensive, the time complexity of all other statistical sampling-based algorithms is $O(n)$, where $n$ is the \emph{sample\_size}, i.e., the number of packets in the sample. As for the online ML algorithms, the time complexity of each \emph{predict()} operation is $O(1)$. The time complexity of each \emph{partial\_fit()} is $O(f^3)$ for the worst online ML algorithm (RLS), and $O(f)$ for the others (SGD and PA), where $f$ is the number of features. Since the number of features we use is relatively small, and it does not change as more packets are processed, we expect the runtime of our framework over different \emph{sample\_size} values to be constant.
		
		\begin{figure}[!tb]
			\centering
			\begin{subfigure}{.95\columnwidth}
				\centering
				\small
				\begin{tabular}{cccc}
\toprule
Mean Sample Size &       GEE &        AE &      UJ2A \\
\midrule
          10,000 &  6.20e-06 &  2.72e-03 &  4.86e-05 \\
          20,000 &  7.69e-06 &  4.06e-03 &  6.99e-05 \\
          30,000 &  9.01e-06 &  5.33e-03 &  8.95e-05 \\
          40,000 &  1.03e-05 &  6.33e-03 &  1.08e-04 \\
          50,000 &  1.14e-05 &  7.44e-03 &  1.34e-04 \\
\bottomrule
\end{tabular}

				\caption{Average execution time in seconds for sampling-based algorithms}
				\label{table:time_estimate}
			\end{subfigure}
			\vspace{5mm}
			
			\begin{subfigure}{.95\columnwidth}
				\centering
				\small
				\begin{tabular}{cccc}
\toprule
Mean Sample Size &       SGD &        PA &       RLS \\
\midrule
          10,000 &  8.81e-06 &  8.52e-06 &  3.55e-06 \\
          20,000 &  7.68e-06 &  8.20e-06 &  3.62e-06 \\
          30,000 &  8.23e-06 &  8.36e-06 &  3.70e-06 \\
          40,000 &  8.15e-06 &  8.36e-06 &  3.53e-06 \\
          50,000 &  8.40e-06 &  9.25e-06 &  3.91e-06 \\
\bottomrule
\end{tabular}

				\caption{Average execution time in seconds of the \emph{predict()} operation for online ML algorithms}
				\label{table:time_predict}
			\end{subfigure}
			
			\vspace{5mm}
			\begin{subfigure}{.95\columnwidth}
				\centering
				\small
				\begin{tabular}{cccc}
\toprule
Mean Sample Size &       SGD &        PA &       RLS \\
\midrule
          10,000 &  2.38e-05 &  3.70e-05 &  2.66e-05 \\
          20,000 &  2.08e-05 &  3.49e-05 &  2.66e-05 \\
          30,000 &  2.09e-05 &  3.49e-05 &  3.06e-05 \\
          40,000 &  2.13e-05 &  3.51e-05 &  2.65e-05 \\
          50,000 &  2.82e-05 &  3.52e-05 &  2.62e-05 \\
\bottomrule
\end{tabular}

				\caption{Average execution time in seconds of the \emph{partial\_fit()} operation for online ML algorithms}
				\label{table:time_fit}
			\end{subfigure}
			
			\caption{Execution time analysis for CAIDA-2016 trace.}
			\label{fig:time_analysis}
		\end{figure}
		
		Figure \ref{fig:time_analysis} shows the execution time of statistical sampling-based algorithms vs. online ML algorithms, over different sample sizes. These measurements are obtained from an Intel Core i7-4500U CPU, with 8GB DDR RAM. The data contains 445 batches taken from the CAIDA-2016 trace, where each batch contains 100,000 packets. For the various online ML algorithms, we used $\{f_i^1, f_i^2, f_i^3, \emph{avg\_pkt\_len}, \emph{syn\_count}\}$ as the feature set. To measure CPU consumption for different \emph{sample\_size} values, we used the following sampling rates: $q = [0.1, 0.2, 0.3, 0.4, 0.5]$. Figure \ref{table:time_estimate} shows the average execution for the sampling-based algorithms,  Figure \ref{table:time_predict} shows the average execution time for the \emph{predict()} operation using several online ML algorithms, and Figure \ref{table:time_fit} shows the average execution time for the \emph{partial\_fit()} operation. As expected, in Figure \ref{table:time_estimate} we can see the linear increase in the CPU consumption with the mean sample size, whereas in Figure \ref{table:time_predict} and \ref{table:time_fit}, CPU consumption is almost constant.

\section{Conclusions} \label{sec:conclusions}
	
		In this work we argued that the problem of current sampling-based algorithms for flow cardinality estimation is their inability to adapt to changes in flow size distribution. Hence, we suggested using online ML framework to add adaptivity to the estimation process, and presented a novel sampling-based adaptive cardinality estimation framework. We analyzed the various possible features, parameters and online ML algorithms for our framework, and proposed the most suitable combination. We tested our framework over real traffic traces, and showed significant improvement in accuracy compared to the best known sampling-based algorithms while using the same amount of computational resources.

\bibliographystyle{abbrv} 
\bibliography{paper}

\begin{thebibliography}{10}

\bibitem{alipourfard2015re}
O.~Alipourfard, M.~Moshref, and M.~Yu.
\newblock Re-evaluating measurement algorithms in software.
\newblock In {\em Proceedings of the 14th ACM Workshop on Hot Topics in
  Networks}. ACM, 2015.

\bibitem{alipourfard2018comparison}
O.~Alipourfard, M.~Moshref, Y.~Zhou, T.~Yang, and M.~Yu.
\newblock A comparison of performance and accuracy of measurement algorithms in
  software.
\newblock In {\em Proceedings of the Symposium on SDN Research}. ACM, 2018.

\bibitem{ben2016heavy}
R.~Ben-Basat, G.~Einziger, R.~Friedman, and Y.~Kassner.
\newblock Heavy hitters in streams and sliding windows.
\newblock In {\em IEEE INFOCOM 2016}.

\bibitem{caidabackbone2016}
{Center for Applied Internet Data Analysis - CAIDA}.
\newblock {The CAIDA UCSD anonymized internet traces 2016 - equinix-chicago}.
\newblock \url{http://www.caida.org/data/passive/passive_2016_dataset.xml}.

\bibitem{caidaddos2007}
{Center for Applied Internet Data Analysis - CAIDA}.
\newblock {The {CAIDA UCSD ``DDoS attack 2007''} dataset}.
\newblock \url{http://www.caida.org/data/passive/ddos-20070804_dataset.xml}.

\bibitem{charikar2000towards}
M.~Charikar, S.~Chaudhuri, R.~Motwani, and V.~Narasayya.
\newblock Towards estimation error guarantees for distinct values.
\newblock In {\em Proceedings of the nineteenth ACM SIGMOD-SIGACT-SIGART}. ACM,
  2000.

\bibitem{cohen2017cardinality}
R.~Cohen, L.~Katzir, and A.~Yehezkel.
\newblock Cardinality estimation meets {Good-Turing}.
\newblock {\em Big Data Research}, 9, 2017.

\bibitem{crammer2006online}
K.~Crammer, O.~Dekel, J.~Keshet, S.~Shalev-Shwartz, and Y.~Singer.
\newblock Online passive-aggressive algorithms.
\newblock {\em Journal of Machine Learning Research}, 7(Mar), 2006.

\bibitem{darpaddos}
{Defense Advanced Research Projects Agency - DARPA}.
\newblock {DARPA scalable network monitoring (SNM) program traffic, PREDICT ID:
  USC-LANDER\textbackslash{}DARPA\_2009\_DDoS\_attack-20091105\textbackslash{}rev4383.}
\newblock \url{http://www.darpa2009.netsec.colostate.edu/}.

\bibitem{deolalikar2016extensive}
V.~Deolalikar and H.~Laffitte.
\newblock Extensive large-scale study of error in samping-based distinct value
  estimators for databases.
\newblock {\em arXiv preprint arXiv:1612.00476}, 2016.

\bibitem{duchi2011adaptive}
J.~Duchi, E.~Hazan, and Y.~Singer.
\newblock Adaptive subgradient methods for online learning and stochastic
  optimization.
\newblock {\em Journal of Machine Learning Research}, 12(Jul):2121--2159, 2011.

\bibitem{duffield2003estimating}
N.~Duffield, C.~Lund, and M.~Thorup.
\newblock Estimating flow distributions from sampled flow statistics.
\newblock In {\em Proceedings of the 2003 conference on Applications,
  Technologies, Architectures, and Protocols for Computer Communications}. ACM,
  2003.

\bibitem{einziger2017constant}
G.~Einziger, M.~C. Luizelli, and E.~Waisbard.
\newblock Constant time weighted frequency estimation for virtual network
  functionalities.
\newblock In {\em ICCCN}. IEEE, 2017.

\bibitem{flajolet2007hyperloglog}
P.~Flajolet, {\'E}.~Fusy, O.~Gandouet, and F.~Meunier.
\newblock Hyperloglog: the analysis of a near-optimal cardinality estimation
  algorithm.
\newblock In {\em AofA: Analysis of Algorithms}. Discrete Mathematics and
  Theoretical Computer Science, 2007.

\bibitem{gale1995good}
W.~A. Gale and G.~Sampson.
\newblock {Good-Turing} frequency estimation without tears.
\newblock {\em Journal of quantitative linguistics}, 2(3), 1995.

\bibitem{gibbons2016distinct}
P.~B. Gibbons.
\newblock Distinct-values estimation over data streams.
\newblock In {\em Data Stream Management}. Springer, 2016.

\bibitem{good1953population}
I.~J. Good.
\newblock The population frequencies of species and the estimation of
  population parameters.
\newblock {\em Biometrika}, 40(3-4), 1953.

\bibitem{grant1987recursive}
I.~H. GRANT.
\newblock Recursive least squares.
\newblock {\em Teaching Statistics}, 9(1), 1987.

\bibitem{haas1998estimating}
P.~J. Haas and L.~Stokes.
\newblock Estimating the number of classes in a finite population.
\newblock {\em Journal of the American Statistical Association}, 93(444), 1998.

\bibitem{hamilton2016quic}
R.~Hamilton, J.~Iyengar, I.~Swett, and A.~Wilk.
\newblock Quic: A {UDP}-based secure and reliable transport for {HTTP/2}.
\newblock {\em IETF, draft-tsvwg-quic-protocol-02}, 2016.

\bibitem{heule2013hyperloglog}
S.~Heule, M.~Nunkesser, and A.~Hall.
\newblock Hyperloglog in practice: algorithmic engineering of a state of the
  art cardinality estimation algorithm.
\newblock In {\em Proceedings of the 16th International Conference on Extending
  Database Technology}. ACM, 2013.

\bibitem{kiefer1952stochastic}
J.~Kiefer and J.~Wolfowitz.
\newblock Stochastic estimation of the maximum of a regression function.
\newblock {\em The Annals of Mathematical Statistics}, 1952.

\bibitem{kingma2014adam}
D.~P. Kingma and J.~Ba.
\newblock Adam: A method for stochastic optimization.
\newblock {\em arXiv preprint arXiv:1412.6980}, 2014.

\bibitem{mouratidis2006continuous}
K.~Mouratidis, S.~Bakiras, and D.~Papadias.
\newblock Continuous monitoring of top-k queries over sliding windows.
\newblock In {\em ACM SIGMOD 2006}.

\bibitem{pedregosa2011scikit}
F.~Pedregosa, G.~Varoquaux, A.~Gramfort, V.~Michel, B.~Thirion, O.~Grisel,
  M.~Blondel, P.~Prettenhofer, R.~Weiss, V.~Dubourg, et~al.
\newblock Scikit-learn: Machine learning in python.
\newblock {\em Journal of Machine Learning Research}, 12(Oct), 2011.

\bibitem{pfaff2015design}
B.~Pfaff, J.~Pettit, T.~Koponen, E.~J. Jackson, A.~Zhou, J.~Rajahalme,
  J.~Gross, A.~Wang, J.~Stringer, P.~Shelar, et~al.
\newblock The design and implementation of {Open vSwitch}.
\newblock In {\em NSDI}, 2015.

\bibitem{tieleman2012lecture}
T.~Tieleman and G.~Hinton.
\newblock Lecture 6.5-rmsprop: Divide the gradient by a running average of its
  recent magnitude.
\newblock {\em COURSERA: Neural networks for machine learning}, 4(2):26--31,
  2012.

\bibitem{uclacsd}
{University of California, Los Angeles, CS department}.
\newblock {UCLA CSD packet traces}.
\newblock \url{https://lasr.cs.ucla.edu/ddos/traces/}.

\end{thebibliography}

\end{document}